\documentclass{emulateapj}
\usepackage{graphicx}
\usepackage{rotating}
\usepackage{pdfpages}
\usepackage{longtable}
\usepackage{epstopdf}
\usepackage{natbib}
\usepackage{amsmath}
\newcommand{\mstar}{$M_*$}
\newcommand{\mhalo}{$M_{\rm halo}$}
\newcommand{\mstareq}{M_*}
\newcommand{\mhaloeq}{M_{\rm halo}}
\newcommand{\mshaloeq}{M^s_{\rm halo}}

\newcommand{\msun}{$M_{\odot}$}
\newcommand{\msuneq}{M_{\odot}}
\newcommand{\rvir}{$r_{vir}$}
\newcommand{\rvireq}{r_{vir}}

\newcommand{\hone}{H~\textsc{i}}

\newcommand{\cfour}{{\rm C}~\textsc{iv}}
\newcommand{\cthree}{C~\textsc{iii}}

\newcommand{\osix}{{\rm O}~\textsc{vi}}
\newcommand{\lya}{Ly$\alpha$~}

\newcommand{\cfourcol}{$N({\text C}~\textsc{iv})$}

\newcommand{\cfourcoleq}{N({\text C}~\textsc{iv})}
\newcommand{\ncfoureq}{\mathcal{N}_{\rm abs}(C~\textsc{iv})}

\newcommand{\beq}{\begin{equation}}
\newcommand{\eeq}{\end{equation}}
\newcommand{\lam}{$\lambda$}

\newcommand{\cmt}{{\rm cm}^{-2}}

\begin{document}
\title{A Deep Search For Faint Galaxies Associated With Very Low-redshift C IV Absorbers: III. The Mass- and Environment-dependent Circumgalactic Medium }
\author{Joseph N. Burchett, Todd M. Tripp, Rongmon Bordoloi, Jessica K. Werk, J. Xavier Prochaska, Jason Tumlinson, C. N. A. Willmer, John O'Meara, Neal Katz}

\begin{abstract}
Using Hubble Space Telescope Cosmic Origins Spectrograph observations of 89 QSO sightlines through the Sloan Digital Sky Survey footprint, we study the relationships between \cfour\ absorption systems and the properties of nearby galaxies as well as large-scale environment.  To maintain sensitivity to very faint galaxies, we restrict our sample to $0.0015 < z < 0.015$, which defines a complete galaxy survey to $L \gtrsim 0.01~L*$ or stellar mass $M_* \gtrsim 10^8~M_{\odot}$. We report two principal findings. First, for galaxies with impact parameter $\rho < 1~\rvireq$, \cfour\ detection strongly depends on the luminosity/stellar mass of the nearby galaxy. \cfour\ is preferentially associated with galaxies with $M_* > 10^{9.5}~M_{\odot}$; lower mass galaxies rarely exhibit significant \cfour\ absorption (covering fraction $f_C = 9^{+12}_{-6}\%$ for 11 galaxies with $M_* < 10^{9.5}~M_{\odot}$).  Second, \cfour\ detection within the $M_* > 10^{9.5}~M_{\odot}$ population depends on environment.  Using a fixed-aperture environmental density metric for galaxies with $\rho <$ 160 kpc at $z < 0.055$, we find that 57$^{+12}_{-13}$\% (8/14) of galaxies in low-density regions (regions with fewer than seven $L > 0.15~L*$ galaxies within 1.5 Mpc) have affiliated \cfour\ absorption; however, none (0/7) of the galaxies in denser regions show \cfour. Similarly, the \cfour\ detection rate is lower for galaxies residing in groups with dark-matter halo masses of $\mhaloeq > 10^{12.5}~\msuneq$. In contrast to \cfour, \textsc{H~i} is pervasive in the CGM without regard to mass or environment. These results indicate that \cfour\ absorbers with log $\cfourcoleq \gtrsim 13.5~\cmt\ $trace the halos of $M_* > 10^{9.5}~M_{\odot}$ galaxies but also reflect larger scale environmental conditions.
\end{abstract}	

\section{Introduction}

The interactions of galaxies over their lifetimes -- with their ambient gaseous surroundings as well as with other galaxies -- is currently a subject of intensive scrutiny.  Galaxies' gaseous halos, or circumgalactic media (CGM), serve as the mediators of processes believed to enable ongoing star-formation and galaxies' eventual quenching and/or transformation.  For example, \textit{feedback}, the combined mechanisms that inject energy and momentum into the interstellar medium (ISM) of galaxies and transport metal-enriched gas expelled by stars or active galactic nuclei (AGN) into their CGM and beyond \citep[e.g., ][]{Oppenheimer:2008mz, Springel:2003lr,Martin:2005qf,Tremonti:2007fk,Bordoloi:2011uq}, is central to the galaxy formation and evolution paradigm. Galaxy formation models that do not include outflows produce galaxies with much higher stellar masses and star-formation rates (SFRs) than those observed \citep{Dave:2011fk,Vogelsberger:2013qf}. Furthermore, these processes are required to enrich the intergalactic medium (IGM) with metals to the observed levels \citep{Cen:2011zl,Oppenheimer:2012qy,Rahmati:2015lr}, and this metal-enriched gas directly traces the products of these feedback processes \citep{Ford:2013lr, Oppenheimer:2006uq,Rubin:2014fk,Veilleux:2005lq}.  

A galaxy's reservoir for star formation is likely to be fed by infalling gas from the IGM \citep{Keres:2005lr,Dekel:2006lr}.  This material, which is metal poor relative to the gas in the disk, has been fervently sought by observers over nearly all wavelengths \citep[e.g.,][]{Kacprzak:2012ul,Ribaudo:2011qf,Rubin:2012fk,Martin:2015kx}, but it is still unclear how galaxies acquire gas and how the physical conditions in the CGM mediate the fueling.  Nevertheless, this diffuse medium is most sensitively probed by absorption line spectroscopy of quasi-stellar objects \citep[QSOs; ][]{Fumagalli:2011qy}, revealing the imprint of neutral hydrogen (\hone) as well as heavy elements in a variety of ionic states.  While \hone\ has been shown to trace the halos of galaxies out to great distances of $\sim$300 kpc and beyond \citep{Morris:1993kq, Tripp:1998kq,Chen:2005kx, Prochaska:2011yq}, the various metal-line transitions probe the physical conditions and enrichment of the gas \citep{Meiring:2013fj,Lehner:2013fj,Werk:2014kx}.

Besides the internal processes mentioned above, galaxy evolution critically depends on the galaxy environment. For over half a century, galaxy clusters have been known to be largely composed of red, early-type galaxies \citep[e.g.,][]{Abell:1965fk} and to be deficient in neutral hydrogen (\hone) \citep{Davies:1973lr}.  As for a physical mechanism relating the cluster environment to the dearth of \hone , \citet{Gunn:1972qy} detailed the process known as `ram-pressure stripping' whereby the hot intracluster medium can remove the gas from an infalling galaxy.  \citet{Dressler:1980qy} showed that the presence of elliptical and S0 galaxies increases with local galaxy density, indicating that the degree by which the corresponding processes impact galaxy morphology increases continuously for denser environments.  Furthermore, \citet{Butcher:1984lr} found that the fraction of blue galaxies in clusters decreases from $z\sim0.45$ to the present and that the spiral galaxies in $z\sim0$ clusters are redder in color than field spirals.  More recently, phenomenology in much less dense regimes has revealed relationships between satellites and central galaxies in groups such that passive satellites preferentially reside around passive centrals and star-forming satellites tend to reside in groups with star-forming centrals \citep{Weinmann:2006fj, Knobel:2015uq}, an effect known as `galactic conformity'.  This phenomenon suggests a deeper, underlying causal relationship between group galaxies and the larger dark matter halos in which they reside. While the question naturally arises whether environmental effects manifest in the CGM, research on the matter \citep{Finn:2016fk,Johnson:2015xy,Johnson:2014rt,Yoon:2013kq,Wakker:2015rf,Tejos:2012lr} is in its infancy.

Both theory and observations suggest that \cfour~absorbers are closely tied to galaxies.  Based on a sample of 14 absorbers at z$<$0.9, \citet{Chen:2001lr} found that strong \cfour~absorbers ($W_{\rm 1548} \gtrsim 200$~m\AA) tend to be found within $\sim$180 kpc of galaxies, regardless of morphology or luminosity.  \citet{Stocke:2013mz} found that the \cfour~systems in their 9-absorber sample lie within $\sim 250$ kpc of galaxies at similar redshifts.  In their simulations, \citet{Ford:2013lr} find that nearly all detectable \cfour~absorption at $z=0.25$ occurs within 300 kpc of galaxies, dropping precipitously at impact parameters $\rho\gtrsim$250 kpc for \mhalo=$10^{11}~\msuneq$ galaxies.  The COS-Dwarfs survey \citep{Bordoloi:2014lr} systematically characterized the \cfour-enriched CGM by targeting a sample of QSO sightline-galaxy pairs with a range of carefully chosen galaxy characteristics, including stellar mass $10^8 M_{\odot} \lesssim M_* \lesssim 10^{10} M_{\odot}$, $\rho<150$ kpc, and specific star formation rate (sSFR) spanning both star-forming and quiescent galaxies.  Among the key COS-Dwarfs results are the following: an anticorrelation between \cfourcol~and impact parameter, with no \cfour~absorbers detected beyond 0.5 virial radii (\rvir), and a tentative correlation between \cfourcol~and sSFR.  \citet{Liang:2014kx} also report a dearth of \cfour~absorbers beyond 0.7 \rvir.    However, both the \citet{Bordoloi:2014lr} and \citet{Liang:2014kx} samples were selected to favor of isolated galaxies, a key difference from this work.  

The work presented here leverages a blind survey of low-redshift \cfour\ absorbers as described in \citet[][hereafter Paper II]{Burchett:2015rf} to investigate the distribution of ionized metal-enriched gas around galaxies down to $\mstareq \sim 10^8 \msuneq$, how this distribution changes with host galaxy mass, and whether the halos traced by \cfour\ show effects of the large-scale environment. The low redshifts of our sample presented in Paper II cover the most recent two billion years of cosmic time and enable deep, high-resolution, multiwavelength studies of the galaxy environments near the absorbers.  At the lowest-redshift end of our distribution ($z\lesssim0.01$), the Sloan Digital Sky Survey\footnote{http://skyserver.sdss3.org} (SDSS) is complete down to very faint galaxies ($L\sim0.01 L*$).  In previous studies, this completeness issue has been perpetually problematic for associating the detected gas with a host galaxy.  At even moderate redshifts of z$\sim$0.2, the angular size scale of modest impact parameters becomes very small and drastically complicates resolving individual galaxies from one another and from the QSO.  By focusing on the lowest redshifts, we exploit the rich galaxy survey data provided by the SDSS and the literature to achieve completeness to faint dwarf galaxies and to begin to assess the relationship between these objects and the absorbing clouds that are readily detectable at all redshifts.  

The paper is organized as follows: The sources of our absorber and galaxy data as well as the methods of deriving measurements from them are described in Section \ref{sec:datameas}.  We present our analyses and results in Section \ref{sec:analysis} and discuss the implications of our study in Section \ref{sec:discussion}. Section \ref{sec:summary} summarizes the paper.  For the reader interested in more information about the large-scale environments analyzed in this work, we offer sightline-centric maps of the galaxies composing them in the Appendix. 

\section{Data and Measurements}
\label{sec:datameas}
The sources of our data and measurements derived from them are described below.  Throughout the paper, we assume nine-year WMAP \citep{Hinshaw:2013zr} cosmology values, with $H_0 = 69.3$, $\Omega_m = 0.29$, and $\Omega_\Lambda = 0.71$.  All stellar mass, halo mass, and star formation rate calculations assume a \citet{Chabrier:2003pd} initial mass function.  Due to the low-redshift nature of our sample, galaxy distances obtained from a pure Hubble flow are very sensitive to the gravitational effects of the Virgo Cluster, the Shapley Supercluster, etc., on measured recession velocities. Therefore, we correct for these effects as a function of object position and redshift using the formalism of \citet{Mould:2000vn}.  These corrections typically result in impact parameters and masses that are within 10\% of their uncorrected values and are well within their uncertainties; however, $\sim5\%$ of objects in our galaxy survey data require distance corrections  $>10\%$ and may affect the mass calculation by up to a factor of $\sim2$, comparable to the uncertainty of our stellar mass calculations.  

\subsection{\cfour~Absorber Sample and Redshift Criteria}
\label{sec:abssample}
The analyses herein most heavily employ the \cfour\ absorber sample from Paper II, which is a blind survey for \cfour\ absorbers within QSO spectra observed with the Hubble Space Telescope Cosmic Origins Spectrograph (HST/COS) in the COS-Halos \citep{Tumlinson:2013cr,Werk:2012qy}, COS-Dwarfs \citep{Bordoloi:2014lr}, and the COS Absorption Survey of Baryon Harbors \citep[CASBaH, ][]{Tripp:2011wd, Meiring:2013fj} programs. Spanning redshifts $0 < z_{\rm abs} < 0.16$, the full sample comprises 42 absorbers that were discovered in 89 sightlines.  Stemming from a blind survey, this absorber sample is free from any galaxy association biases inherent in targeted galaxy/absorber studies and indeed does not impose \textit{a priori} that a given absorber is associated with any galaxy at all.

For our analyses in this paper, the existing galaxy survey data around these absorber sightlines (see Section \ref{sec:galsurveys}) do not have sufficient spectroscopic completeness for galaxies over the full $z=0-0.16$ range where \cfour\ is covered by the high-resolution COS gratings. Therefore, the two primary but complementary analyses presented in this paper adopt redshift limits commensurate with the requirements of each: 

(1) A defining objective of this series of papers is to surpass the sensitivity of previous surveys to dwarf galaxies potentially associated with \cfour\ absorbers and assess the connection of these faint dwarfs, undetected at higher redshifts, to the absorbers.  Thus, we first examine absorber/galaxy associations as a function of galaxy mass, where the mass range spans down to an order of magnitude below the lowest mass where circumgalactic \cfour\ has been previously characterized by galaxy-selected studies \citep{Bordoloi:2014lr, Liang:2014kx}.  The median galaxy masses of the COS-Dwarfs \citep{Bordoloi:2014lr} and \citet{Liang:2014kx} samples are $2 \times 10^9~\msuneq$ and $5 \times 10^9~\msuneq$, respectively.  Using the M$_*$-absolute $r$-band magnitude ($\mathcal{M}_r$) relation derived by \citet{Liang:2014kx} from the NASA-Sloan Atlas\footnote{http://www.nsatlas.org}, a stellar mass of $2 \times 10^8~\msuneq$ corresponds to $\mathcal{M}_r = -16.6$, approximately $L_r \sim 0.01~L*$ \citep{Blanton:2003kx}.  Given its apparent magnitude limit, the SDSS should be complete to galaxies of this mass at $z \lesssim 0.015$.  Therefore, our absorber-galaxy mass analysis adopts a redshift limit of $z \lesssim 0.015$.

(2) Following on the results of (1) where we find an increased \cfour\ detection rate for $\mstareq > 10^{9.5} \msuneq$ galaxies over less massive galaxies, we then investigate the dependence of CGM absorption on the larger environments of these galaxies.   Because we select the CGM/environmental analysis sample on $\mstareq > 10^{9.5} \msuneq$ galaxies (corresponding to $\mathcal{M}_r \lesssim -19$), we expect spectroscopic completeness to these galaxies over the full redshift range of the NASA-Sloan Atlas.  Therefore, we adopt the upper redshift limit of the atlas, $z=0.055$, minus a small buffer to mitigate edge effects, for this analysis, including absorbers at $z<0.052$ and their measurements from COS-Dwarfs \citep{Bordoloi:2014lr} and the blindly-detected sample of Paper II \citep{Burchett:2015rf}.  

The column densities of detected absorbers were obtained via Voigt profile fitting.  The upper and lower limits reported here were measured by the apparent optical depth method \citep{Savage:1991vn} using the error vector output from the COS data reduction pipeline and assuming a velocity window of $\pm$ 50 km/s of the corresponding galaxy redshifts.  Covering fractions reported for galaxies/absorbers from this work assume 3-$\sigma$ detection thresholds of log $\cfourcoleq > 13.5~\cmt$ and log N(\hone)$~> 13.0~\cmt$.

\subsection{Optical Galaxy Surveys}
\label{sec:galsurveys}
As all but one of the QSOs observed in our dataset (Paper II) fall within the SDSS footprint, this study employs the SDSS extensively, specifically SDSS Data Release 12 \citep{Alam:2015kq} and the NASA-Sloan Atlas, for both spectroscopy and photometry.  We also exploit the NASA/IPAC Extragalactic Database\footnote{http://ned.ipac.caltech.edu; The NASA/IPAC Extragalactic Database (NED) is operated by the Jet Propulsion Laboratory, California Institute of Technology, under contract with the National Aeronautics and Space Administration.}, which houses catalogs from several large galaxy surveys, such as the Third Reference Catalogue of Bright Galaxies \citep[RC3;][]{Vaucouleurs:1991rt,Corwin:1994vn} and the Two-degree-Field Galaxy Redshift Survey \citep[2dFGRS;][]{Colless:2001ys}.  The NASA-Sloan Atlas is a catalog compiling newly remeasured SDSS and GALEX photometry along with spectral line measurements using an improved calibration of the SDSS spectra.  The core sample for this catalog stems from SDSS DR8 and adds objects missed or excluded by SDSS but spectroscopically measured by the CfA Redshift Survey \citep{Huchra:1995bh}, 2dFGRS, etc.  According to \citet{Strauss:2002fk}, the SDSS spectroscopic survey is complete down to $m_r \sim 17.7$.  All magnitudes expressed herein are derived from the model-fit magnitudes from the NASA-Sloan Atlas, which calculate total magnitudes of galaxies by Sersic profile fits, or the `modelMag' values, those deemed the better fit between de Vaucoleurs and exponential profiles, directly from the SDSS database for objects not provided by NASA-Sloan.  

At the low redshifts ($z < 0.015$) primarily targeted in this work, even low-mass galaxies appear quite extended on the sky and a number of issues occur within the SDSS databases.  For the analyses herein, we primarily employ the NASA-Sloan Atlas instead of the `stock' SDSS databases.  However, we also produced our own SDSS catalog to corroborate results obtained using the NASA-Sloan Atlas and specifically address certain linking issues within SDSS: First, extended objects may have received multiple spectroscopic measurements and thus may be duplicated in the spectroscopic object (`SpecObj') tables.  Second, we found numerous cases of galaxy spectra linked to incorrect galaxy photometric objects, thus yielding clearly incorrect redshifts retrieved through queries to the `Galaxy' table of the SDSS database.  Third, spectroscopic data are primarily keyed to photometric objects via the `FluxObjID' field in the `SpecObj' table, and these linked photometric objects are sometimes `Child' objects that do not reflect the full integrated photometry of the galaxy.  Finally, certain nearby galaxies were not targeted by the SDSS.   Therefore, we produced a new spectroscopic galaxy catalog by (1) querying the SDSS `SpecObj' table for all galaxies within 7 degrees of the QSO sightlines (corresponding to 1 Mpc projected on the sky at $z=0.002$), (2) running an internal crossmatch using the STILTS software package \citep{Taylor:2006yu} to find duplicate objects, (3) identifying objects with suspected incorrectly linked photometry ($m_r > 19$ when the nominal SDSS completeness limit is $m_r = 17.7$) and requerying the SDSS `PhotoObj' table for brighter objects that may be the true `parent' photometric object within $~10$ arcsec, and (4) incorporating galaxies from RC3, which complements the SDSS by including the bright objects SDSS may not have targeted (approximately 650 galaxies over our full sample).

The analysis presented in Section \ref{sec:envirociv} utilizes the group halo catalog of \citet{Yang:2007kl} (specifically their version updated for SDSS DR7), which includes the results of a group finding algorithm to identify galaxy groups within the SDSS at $z>0.01$.  The data products\footnote{http://gax.shao.ac.cn/data/Group.html} include group centroids, halo masses, and the identified galaxy members of each group, which we use to assign halo masses and virial radii to the larger dark matter halos in which our absorbers/galaxies reside.

\subsection{Stellar Masses, Halo Masses, and Virial Radii}
\label{sec:massesrvir}
Our galaxy stellar masses (\mstar) are calculated using SDSS photometry and the \textsc{kcorrect} software, which fits galaxy spectral templates to broadband photometric data \citep {Blanton:2007ys}.  The NASA-Sloan Atlas provides stellar masses output from \textsc{kcorrect} using the improved photometry, and we used the software to also calculate stellar masses for galaxies not included in the atlas.  

Our analyses consider galaxy halos on two scales: individual galaxies and groups/clusters.  Our first objective is to characterize the galaxies associated with \cfour\ absorbers at low redshift, and the extent of a galaxy's dark matter halo provides a natural scale for associating an absorbing gas cloud with the galaxy.  We adopt $r_{200}$, the radius within which the average mass density is 200 times the critical density of the Universe, as the virial radius (\rvir) of a galaxy halo. To estimate the total halo masses (\mhalo) and virial radii of the galaxies around QSO sightlines, we primarily employ the redshift-dependent stellar mass/halo mass relation of \citet{Moster:2013lr} and the calculation described by \citet{Burchett:2013qy}.  However, similar to \citet{Prochaska:2011yq} and \citet{Stocke:2013mz}, both of whom analyzed galaxy/absorber associations with respect to \rvir, we employ an \mstar-\mhalo relation that diverges from that obtained through pure halo abundance matching for massive galaxies.  \citet{Moster:2013lr} parameterize the \mstar/\mhalo\ ratio using a double power-law form with slopes above and below a characteristic mass:

\beq
\label{eqn:pureHAM}
\frac{\mstareq}{\mhaloeq} = 2N \left[  \left(\frac{\mhaloeq}{M_1}\right)^{-\beta} + \left(\frac{\mhaloeq}{M_1}\right)^{\gamma} \right]^{-1}
\eeq

\noindent where $N$ is the normalization and $M_1$ is the characteristic mass above and below which the behavior of \mstar/\mhalo\ follows the slopes $\beta$ and $\gamma$.  This form is motivated by the deviation of the galaxy luminosity function from the halo mass function at low and high masses \citep{Yang:2003yq}, and it is assumed that the galaxy populations within halos below a certain \mhalo, on average, are dominated by the central galaxy.  Higher-mass halos will be increasingly populated by satellite galaxies in number, and by comparing the conditional mass function of centrals with the full stellar mass function of halos at varying \mhalo, \citet{Moster:2010rt} find that the satellite contribution becomes increasingly important above log $\mhaloeq = 12.0~\msuneq$.  Therefore, a stellar-to-halo mass conversion from halo abundance matching for $L>L_*$ galaxies (assuming the galaxy is a central) provides \mhalo\ values appropriate for group and cluster scales; we employ this conversion, using the Equation \ref{eqn:pureHAM} parameter values from \citet{Moster:2013lr}, for \mhalo\ and \rvir\ in the galaxy environment/absorber analysis.

Aiming to produce a scaling more suited to individual galaxies across a wide dynamic range in stellar mass, we modify the \mstar-\mhalo\ conversion as follows.   In the $\mhaloeq > M_1$ regime (log $M_1 = 11.590$ at $z=0$), the \mstar/\mhalo\ ratio decreases with increasing \mhalo, and halos become dominated by satellite galaxies at log $\mhaloeq > 12.0~\msuneq$.  Thus, we effectively truncate the contribution of the second term by fixing it at \mhalo = $M_1$, which is firmly within the central-dominated regime but where the contribution of this term is non-negligible.  Equation \ref{eqn:pureHAM} becomes

\beq
\label{eqn:HAMsingle}
\frac{\mstareq}{\mshaloeq} (z=0) = 0.0702 \left[  \left(\frac{\mshaloeq}{10^{11.590}}\right)^{-1.376} + 1 \right]^{-1}
\eeq

\noindent where we have also substituted the $z=0$ values for $M_1$, $\beta$, and $\gamma$.  The halo masses used to associate individual galaxies with absorbers via the virial radius were then obtained by numerically solving the analog of Equation \ref{eqn:HAMsingle} at the redshift and stellar mass of each galaxy in the database using the appropriate parameters from \citet{Moster:2013lr}.

\subsection{Star Formation Rates}
\label{sec:SFR}

Throughout the paper, figures are color coded to reflect whether the galaxies plotted are star forming (blue) or quiescent (red).  This distinction was made based on star formation rates (SFRs) calculated from either the GALEX FUV and NUV photometry provided by the NASA-Sloan Atlas or the H$\alpha$ and H$\beta$ spectral line measurements from SDSS spectroscopy. Because  good quality GALEX data were not available for many galaxies, mostly due to problems with the FUV band, the Balmer line measurements provided a secondary means to measure or place limits on the SFRs.  The primary disadvantage for these spectroscopically derived values is the need for flux correction due to the small angular size of the SDSS spectrograph fibers compared to the full projected size of the galaxies targeted.  

SFRs using the GALEX data were calculated using Equations (5), (7), and (8) of \citet{Salim:2007lr}. The SFRs from Balmer line measurements were calculated using Equation (4) of \citet{Brinchmann:2004lr}, where the H$\alpha$ luminosity was corrected for dust attenuation using the intrinsic H$\alpha$-to-H$\beta$ luminosity assuming Case B recombination and $10^4$ K gas from \citet{Osterbrock:2006fk}, and $k$(H$\alpha$)-$k$(H$\beta$) was calculated from the extinction law of \citet{Cardelli:1989lr}.  The Balmer line-derived SFRs were fiber corrected by scaling the flux of each line by the ratio of the total `model' NUV flux, resulting from the photometric fit of the entire galaxy, to the `fiber' flux, which has an aperture equal to the size of the SDSS fiber.  If the NUV data were not available, we used the $u$-band fluxes in an attempt to capture, as much as possible, stellar continuum from the youngest populations.  A comparison of the total-to-fiber flux ratios between the NUV and SDSS $u$-band, where the NUV data were available, revealed that the $u$-band ratios were systematically lower than the NUV ratios. This difference in flux ratios is likely driven by two effects: (1) GALEX better captures more diffuse star formation occurring in the outer regions of galaxies and (2) the $u$-band flux has a greater contribution from older stars and is more centrally concentrated.  For comparison, the $r$-band in turn yields lower total-to-fiber flux ratios than the $u$-band, supporting this second point. We thus considered the Balmer-line SFRs using $u$-band flux corrections as lower limits.  In cases where the FUV and NUV data were not available and the H$\alpha$ line was not detected at 3$\sigma$ significance, we adopted a 3-$\sigma$ upper limit from the H$\alpha$ line measurement uncertainty scaled for the fiber correction.

With these SFRs in hand, we set a threshold separating passive and star-forming galaxies based on the the bimodality in the locus of specific star formation rate (sSSFR=SFR/\mstar) vs. \mstar\ for the entire NASA-Sloan Atlas at log (sSFR/yr$^{-1}$) = -10.75.  Galaxies with sSFR above and below this threshold are represented in blue and red, respectively, in figures throughout the paper.

\begin{figure*}[t!]
\centering
\includegraphics[width=0.8\paperwidth]{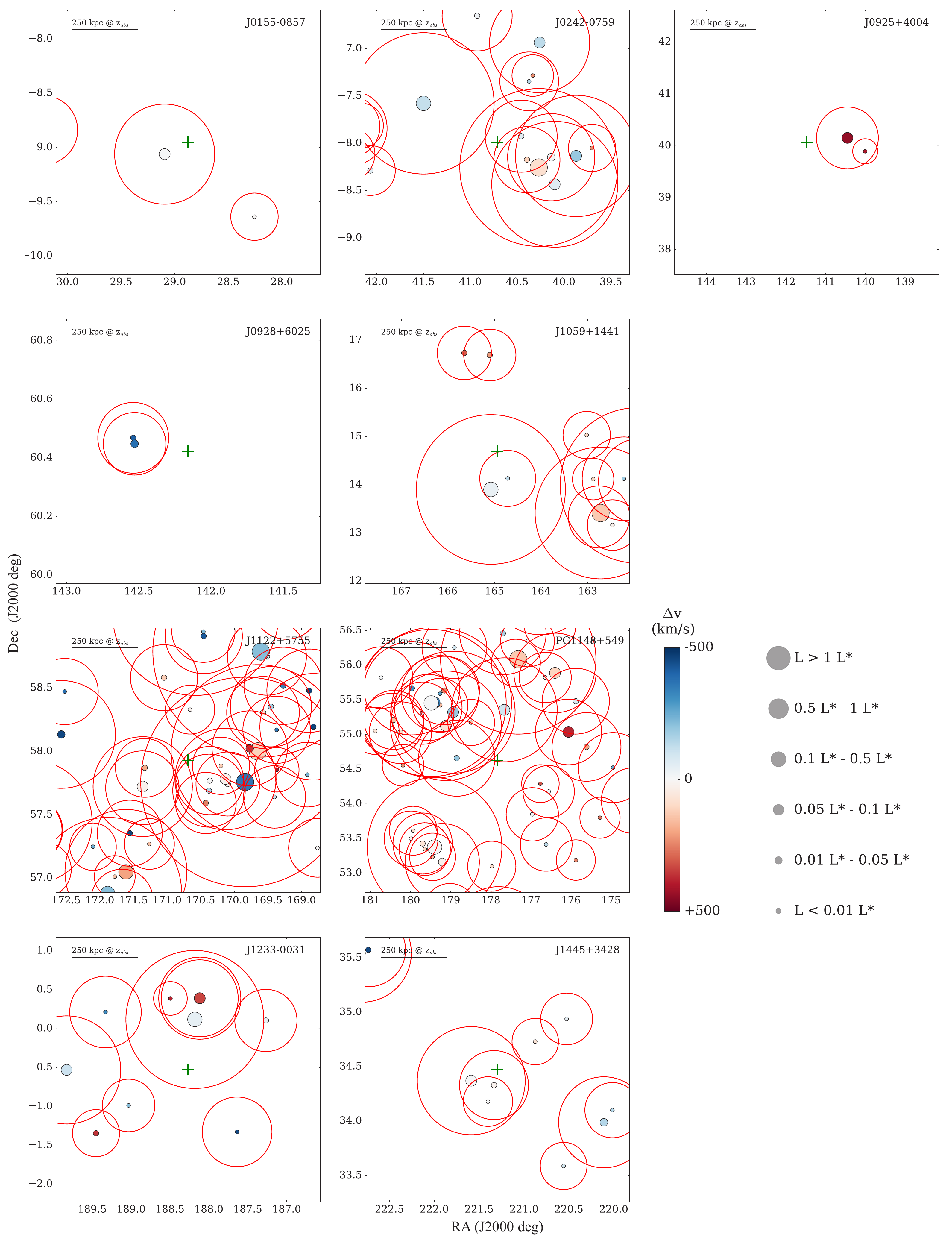}
\caption{Galaxies around QSO sightlines within $\pm$ 500 km/s of the $z<0.015$ absorber subsample.  Following the legend and scalebar on the right side, the marker size indicates galaxy $r-$band luminosity, the marker color indicates the separation in radial velocity from each absorber, and the red circles mark the virial radius of each galaxy.  The abbreviated field name is printed in the upper-right corner of each panel.  A physical scale	 is given in each panel.  Note that all galaxies in the J0925+4004 frame have $v_{\rm gal} - v_{\rm abs} > 700$ km/s and thus lie at much larger impact parameters than that indicated by the reference scale.}
\label{fig:absmaps}
\end{figure*}

\section{Analysis and Results}
\label{sec:analysis}
We now investigate the connections between \cfour~absorbers and nearby galaxies.  In particular, we pose two overarching questions: 1) How is the material traced by \cfour\ absorption distributed relative to nearby galaxies in terms of proximity to the associated galaxies and as a function of galaxy mass?  2) How, if at all, is the CGM affected by the environment in which a host galaxy resides?  


\subsection{Absorption profiles and the dependence of \cfour\ absorption on galaxy mass}
\label{sec:indivgals}
Here, we complement the galaxy-selected approach of COS-Dwarfs with a blindly detected sample of \cfour~absorbers (Paper II) for which we search the SDSS for galaxy counterparts.  As stated in section \ref{sec:abssample}, these analyses employ a $z \leq 0.015$ galaxy/absorber subsample that includes detected \cfour\ absorbers as well as upper limits on the \cfour~absorption at redshifts where galaxies fall within 500 kpc of a QSO sightline.  For comparison, we also include similar relations for \hone~\lya. As employed by COS-Halos \citep{Tumlinson:2013cr} and COS-Dwarfs, we adopt a search window of $|\Delta v| < 600$ km/s for velocity separations between galaxies and absorbers.

First, we describe our selection methods for associating galaxies to absorbers; then, we present the resulting column density-impact parameter profiles and \cfour\ detection statistics as a function of associated galaxy mass.

\subsubsection{Selecting Absorber-associated Galaxies}
\label{sec:selection}
Nine \cfour~absorbers from our sample meet the $z~<~0.015$ criterion, and we consider associated galaxies based on two selection methods: we identify the galaxy with the smallest projected separation from the sightline in terms of (1) proper distance (kpc) and 2) galaxy virial radius ($\rho/$\rvir).  We present maps of the galaxy fields within proper distances of 500 kpc surrounding each $0.0015 < z_{abs} < 0.015$ absorber in Figure \ref{fig:absmaps}.  Note that these maps contain \textit{all} galaxies with available spectroscopic redshifts and have no imposed minimum luminosity.  Each galaxy's virial radius is represented by a red circle. As shown in the maps, the \cfour~systems largely occur in galaxy environments that are sufficiently populated to preclude unambiguous selection of a `host galaxy', and we proceed by selecting as `associated' the most proximal galaxies to each sightline using  the two metrics mentioned above and examining the resulting distributions of absorption (or lack thereof) with respect to these two selections. With the exception of one absorber ($z_{\rm abs} = 0.00261$ in the sightline of QSO J0925+4004, but see the discussion below), we find galaxies within impact parameters of 200 kpc for all absorbers in this subsample.  Certain \cfour~absorbers are located outside of the virial radii of all nearby galaxies, but only moderately beyond the virial radii.

These maps underscore the difficulty of associating absorbers with individual galaxies.  Only the field of J0155-0857 presents a fairly unambiguous association between an absorber and galaxy regardless of selection method. The J0242-0759, J1059+1441, J1122+5755, 
and J1445+3428 fields contain absorbers that appear to fall within the virial radii of multiple galaxies.  J0925+4004, J0928+6025, and PG1148+549 contain absorbers outside the virial radius of any galaxy at the same redshift.  Although several fields depicted in Figure \ref{fig:absmaps} show an abundance of galaxies near the QSO sightline (due to the high completeness to faint galaxies), the two selection schemes we have employed capture the essential absorber/galaxy trends and enable comparison to previous studies.

The absorber in J0925+4004 does not appear to have a galaxy within 300 kpc or 1.7 \rvir~ even if we expand the velocity search window to $|\Delta v| < 800$ km/s (note that the galaxies depicted in Figure \ref{fig:absmaps} for this absorber all exceed velocity differences of 600 km/s). Another galaxy, SDSS J092422.92+400527.1, occurs at 141 kpc and 1.2 \rvir~if the velocity window is widened to 850 km/s.  However, this low-surface brightness galaxy's SDSS spectrum is quite noisy, and if its redshift is equal to that of the absorber, the impact parameter decreases to 56 kpc.  If the SDSS redshift is accurate, this last absorber presents a similar issue to that presented by \citet{Burchett:2013qy}, where a candidate host galaxy may have a smaller impact parameter if the gas is moving at a very high velocity relative to the galaxy, e.g., a high-velocity outflow.  Such outflow speeds are certainly plausible and have in fact been observed unambiguously in `down-the-barrel' studies where the ejecting galaxy itself was used as a background source to study the outflowing gas in absorption \citep[e.g., ][]{Rubin:2014fk, Tremonti:2007fk}.  However, using distant QSOs as the background continuum source probes circumgalactic gas that is transversely separated in projection from possible host galaxies and introduces ambiguity in cases where large velocity offsets may not indicate large line-of-sight Hubble flow distances.  We defer a deeper analysis of the velocity offset/impact parameter degeneracy to future galaxy surveys that will improve statistics while maintaining a sensitivity to faint dwarfs (Burchett et al., in preparation).  

\begin{deluxetable*}{cccccccccc} 
\centering
\tabletypesize{\scriptsize} 
\tablewidth{0pt} 
\tablecaption{Closest SDSS galaxies to  z$<$0.015 \ion{C}{4} absorption systems, selected by proper distance.} 
\tablehead{  QSO &  $z_{abs}$ &   Galaxy &  $\alpha_{gal}$ (J2000) &  $\delta_{gal}$ (J2000)  &  $z_{gal}$ &  $\rho$  &  $\delta$v &  log $M_*$ &  $\rho/r_{vir}$ \\ 
 \ & \ & \ & \multicolumn{2}{c}{(degrees)} & \ & (kpc) & (km/s) & ($M_{\odot}$) & \ }
\startdata 
J0155-0857 & 0.00547 & NGC 0755 & 29.09370 & -9.06231 & 0.00547 & 98 & 0 & 9.67 & 0.53 \\  
J0242-0759 & 0.00477 & SDSS J024149.95-075530.0 & 40.45766 & -7.92484 & 0.00458 & 88 & 57 & 8.65 & 0.69 \\  
J0925+4004 & 0.00261 & NGC 2844 & 140.45005 & 40.15125 & 0.00496 & 369 & -701 & 10.26 & 1.50 \\  
J0928+6025 & 0.01494 & SBS 0926+606A & 142.52826 & 60.44781 & 0.01366 & 201 & 378 & 8.18 & 1.83 \\  
J1059+1441 & 0.00242 & SDSS J105851.95+140748.2 & 164.71646 & 14.13006 & 0.00199 & 65 & 128 & 7.15 & 0.83 \\  
J1122+5755 & 0.00640 & UGC 06369 & 170.36413 & 57.76851 & 0.00640 & 139 & 0 & 8.81 & 1.02 \\  
J1233-0031 & 0.00392 & SDSS J123718.74+001248.0 & 189.32907 & 0.21289 & 0.00282 & 137 & 326 & 7.18 & 1.72 \\  
J1445+3428 & 0.00549 & SDSS J144520.23+341948.1 & 221.33415 & 34.33011 & 0.00556 & 80 & -20 & 8.93 & 0.56 \\  
PG1148+549 & 0.00349 & UGC 06894 & 178.84798 & 54.65728 & 0.00284 & 185 & 193 & 8.67 & 1.42 \\  
\enddata 
\label{tab:kpcgals} 
\end{deluxetable*} 

\begin{deluxetable*}{cccccccccc} 
\centering
\tabletypesize{\scriptsize} 
\tablewidth{0pt} 
\tablecaption{Closest SDSS galaxies to  z$<$0.015 \ion{C}{4} absorption systems, selected by virial radius.} 
\tablehead{  QSO &  $z_{abs}$ &   Galaxy &  $\alpha_{gal}$ (J2000) &  $\delta_{gal}$ (J2000) &  $z_{gal}$ &  $\rho$  &  $\delta$v &  log $M_*$ &  $\rho/r_{vir}$ \\ 
 \ & \ & \ & \multicolumn{2}{c}{(degrees)} & \ & (kpc) & (km/s) & ($M_{\odot}$) & \ }
\startdata 
J0155-0857 & 0.00547 & NGC 0755 & 29.09370 & -9.06231 & 0.00547 & 98 & 0 & 9.67 & 0.53 \\  
J0242-0759 & 0.00477 & NGC 1052 & 40.27000 & -8.25578 & 0.00504 & 192 & -79 & 10.74 & 0.45 \\  
J0925+4004 & 0.00261 & NGC 2844 & 140.45005 & 40.15125 & 0.00496 & 369 & -701 & 10.26 & 1.50 \\  
J0928+6025 & 0.01494 & SBS 0926+607 & 142.53767 & 60.46827 & 0.01355 & 209 & 411 & 8.53 & 1.70 \\  
J1059+1441 & 0.00242 & NGC 3489 & 165.07691 & 13.90107 & 0.00230 & 106 & 35 & 10.23 & 0.44 \\  
J1122+5755 & 0.00640 & NGC 3613 & 169.65041 & 58.00007 & 0.00678 & 343 & -115 & 11.09 & 0.42 \\  
J1233-0031 & 0.00392 & NGC 4517 & 188.17980 & 0.11716 & 0.00376 & 206 & 45 & 10.43 & 0.73 \\  
J1445+3428 & 0.00549 & SDSS J144520.23+341948.1 & 221.33415 & 34.33011 & 0.00556 & 80 & -20 & 8.93 & 0.56 \\  
PG1148+549 & 0.00349 & NGC 3913 & 177.66224 & 55.35397 & 0.00317 & 255 & 96 & 9.86 & 1.27 \\  
\enddata 
\label{tab:rvirgals} 
\end{deluxetable*}

Selecting a galaxy to be associated with an absorber based on the closest transverse proper distance is the scheme employed by most comparable surveys to date. In many ways, this criterion is the most intuitive and captures a natural suspect for the source of the gas based on pure proximity.  Also, this scheme is relatively free of the assumptions inherent in an assignment based on galaxy properties, such as the virial radius.  However, as a survey becomes sensitive to fainter galaxies, the luminosity function \citep{Schechter:1976yu} indicates that the number density of galaxies will increase as $\phi \propto L^{\alpha}$, where recent estimates place $\alpha \sim -1.3$ for the full star-forming and quiescent population \citep{Loveday:2012jk,Willmer:2006lr}.  While this increases the likelihood of detecting a galaxy closer to the sightline, the question remains whether the fainter galaxy is the true source of the gas.

Galaxies potentially associated with absorbers may also be selected based on galaxy virial radii. The galaxy's virial radius provides an estimate of a galaxy halo's extent, insomuch as gas within the virial radius is consistent with being bound to the galaxy provided its velocity separation is also less than the galaxy's escape velocity.  Indeed, previous studies have reported that the column density of \cfour\ absorbers anticorrelates with impact parameter relative to the host galaxy virial radius \citep[$\rho/r_{\rm vir}$;][]{Bordoloi:2014lr, Liang:2014kx}.

To verify the advertised completeness of our galaxy data, we investigated what galaxies down to the desired luminosity limit may have been missed by the publicly available redshift surveys between the QSO sightlines and the associated galaxies we identified.  We scaled the galaxies around each sightline from the SDSS photometry that did not have SDSS spectroscopic redshifts to their hypothetical absolute r-band magnitudes, assuming they were all at the redshift of the absorber.  We then estimated the galaxy luminosities relative to L* using the r-band absolute magnitude $M*_r = -20.44 - 5~\rm{log}~h$ \citep{Blanton:2003kx} and the SDSS dereddened r-band model-derived magnitudes, which use extinction values from the \citet{Schlegel:1998kq} dust maps.  As a result, we find that our galaxy redshift data (SDSS, etc.) are 100\% complete to $0.01~L*$ in the region of interest around 8 sightlines in this subsample and 100\% complete to $0.025~L*$ in the one remaining sightline ($z=0.01494$ in the J0928+6025 sightline).  These luminosities are commensurate with the survey goals outlined in Section \ref{sec:abssample}.

\begin{figure*}[t!]
\centering
\includegraphics[width=0.9\paperwidth]{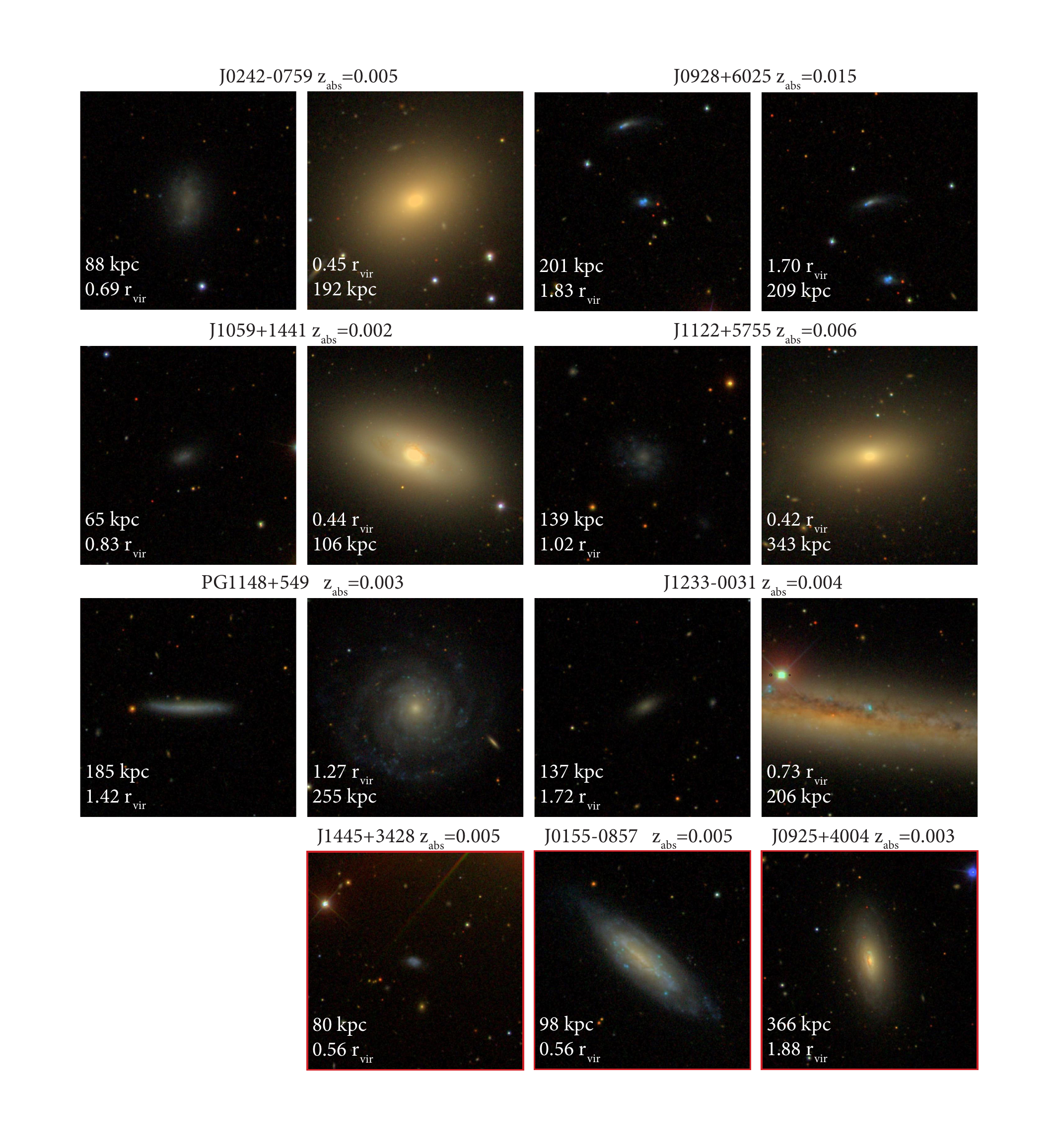}
\caption{SDSS false color images \citep[from $g$, $r$, and $i$ bands;][]{Lupton:2004yu} of galaxies selected to be associated with our $z<0.015$ \cfour\ absorber subsample.  We selected associated galaxies by both the smallest proper distance impact parameter and impact parameter relative to the virial radii of the galaxies in the field.  Except for the three bottom panels marked with red boxes, panels are paired for each absorber (labeled above) where the left panel shows the galaxy selected by proper distance and the right shows the galaxy selected by virial radius.  Each panel depicts a projected area on the sky approximately 3.5 arcminutes per side.}
\label{fig:compareselections}
\end{figure*}

The absorber-galaxy associations were then drawn as follows: (1) We used  the traditional method of assigning an absorber to the galaxy with the  smallest impact parameter and velocity separation from the absorber.   We refer to the sample selected in this way as \textit{proper-distance  selected}. (2) Alternatively, we assigned the absorber to the galaxy at the absorber redshift that is closest in terms of the fraction of  its virial radius ($\rho$/\rvir).  This alternative selection  produces our \textit{virial-radius selected} sample. The resulting galaxies  selected by these two methods are shown in Tables 1 and 2,  respectively.  SDSS composite images of the selected galaxies are shown in Figure \ref{fig:compareselections}, where a pair of panels is shown for each absorber in the 6/9 cases where the two selection methods choose different galaxies, and single panels are shown for the three cases where the selection methods choose the same galaxy (marked with red borders). 

In addition to searching for galaxies associated with detected absorbers, to quantify the cases where no absorption was detected but galaxies lie near the sightline, we also searched the SDSS for $z_{gal} < 0.015$ galaxies at impact parameters within 500 kpc of \textit{any} of the 89 QSO sightlines described in Section \ref{sec:abssample}. We further filtered the galaxies found in this search by iterating through each galaxy to identify the galaxy with the smallest impact parameter, both in proper distance and virial radius, of any galaxy within a redshift range of $\delta z_{gal} < 0.003$, or $\sim 900$ km/s.  This velocity range was chosen to exceed that for which we matched \textit{detected} absorbers and galaxies (600 km/s) and mitigate confusion between galaxies that may lie at similar redshifts to one another due to peculiar motion but fall at similar line-of-sight cosmological distances. Also, galaxies targeted by the COS-Dwarfs survey were rejected for this stage of the analysis to further ensure that our sample is blind in both absorber and galaxy selection (although we include them later for the environmental analysis where their inclusion does not introduce bias).  

\subsubsection{\cfour~column density and galaxy impact parameter}
\label{sec:c4colprofiles}
\begin{figure}[t!]
\includegraphics[width=.52\textwidth]{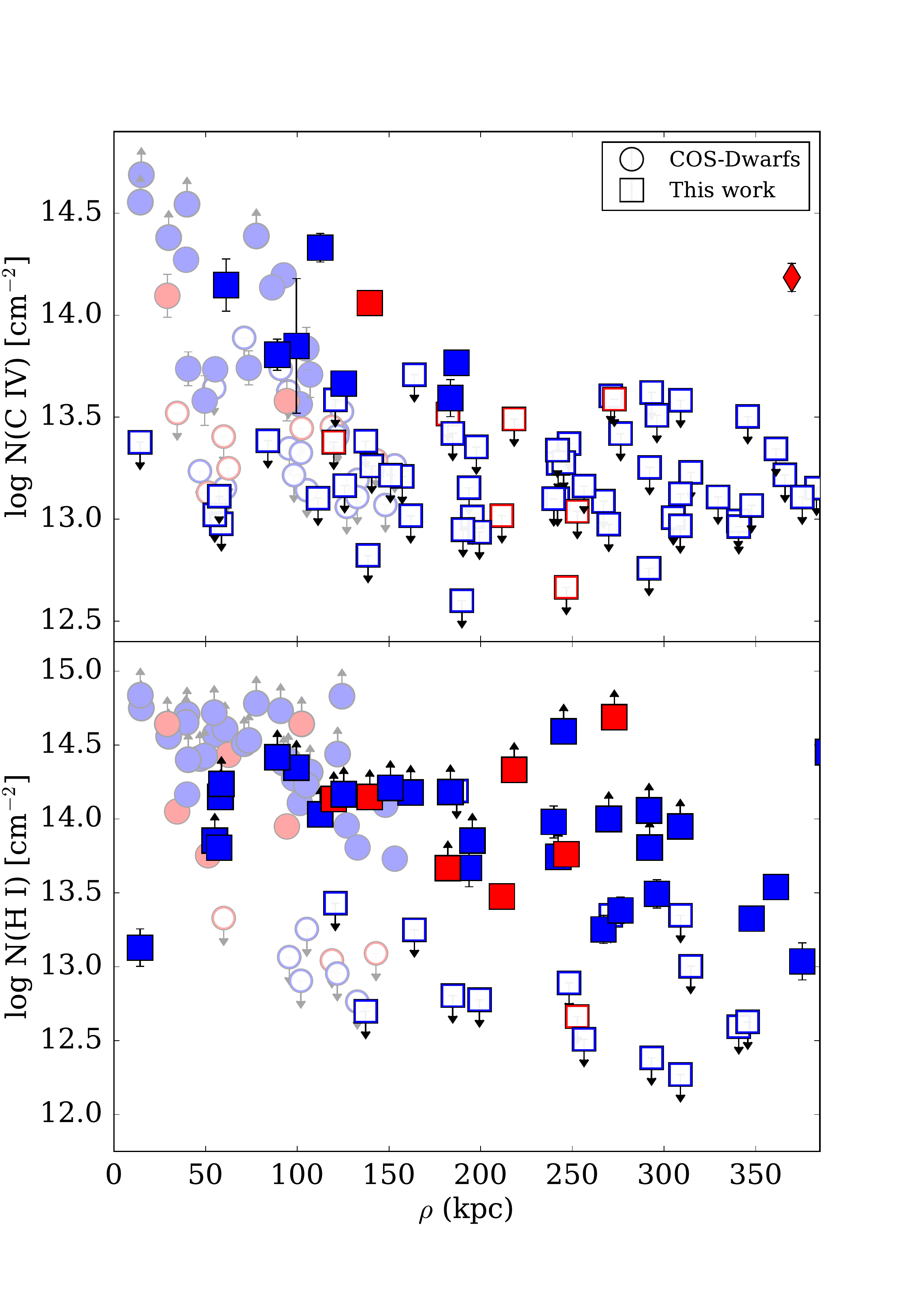}
\caption{The \cfour\ (top) and \hone\ (bottom) column density profiles for our low-z absorber/galaxy sample (squares).   Associated galaxies were selected by proper distance proximity using a galaxy magnitude limit of $\mathcal{M}_r \leq -16.4$.  The galaxy/absorber pair indicated with a diamond in the top panel has the greatest velocity separation of all pairs plotted, with $\delta v >600$ km/s. Red and blue symbols denote passive and star-forming galaxies, respectively. The open symbols with downward arrows correspond to  3-$\sigma$ upper limits on the absorption; filled symbols with upward arrows indicate lower limits measured using the apparent optical depth on saturated lines. }
\label{fig:kpcimpact}
\end{figure}

\begin{figure}[t!]
\centering
\includegraphics[width=.52\textwidth]{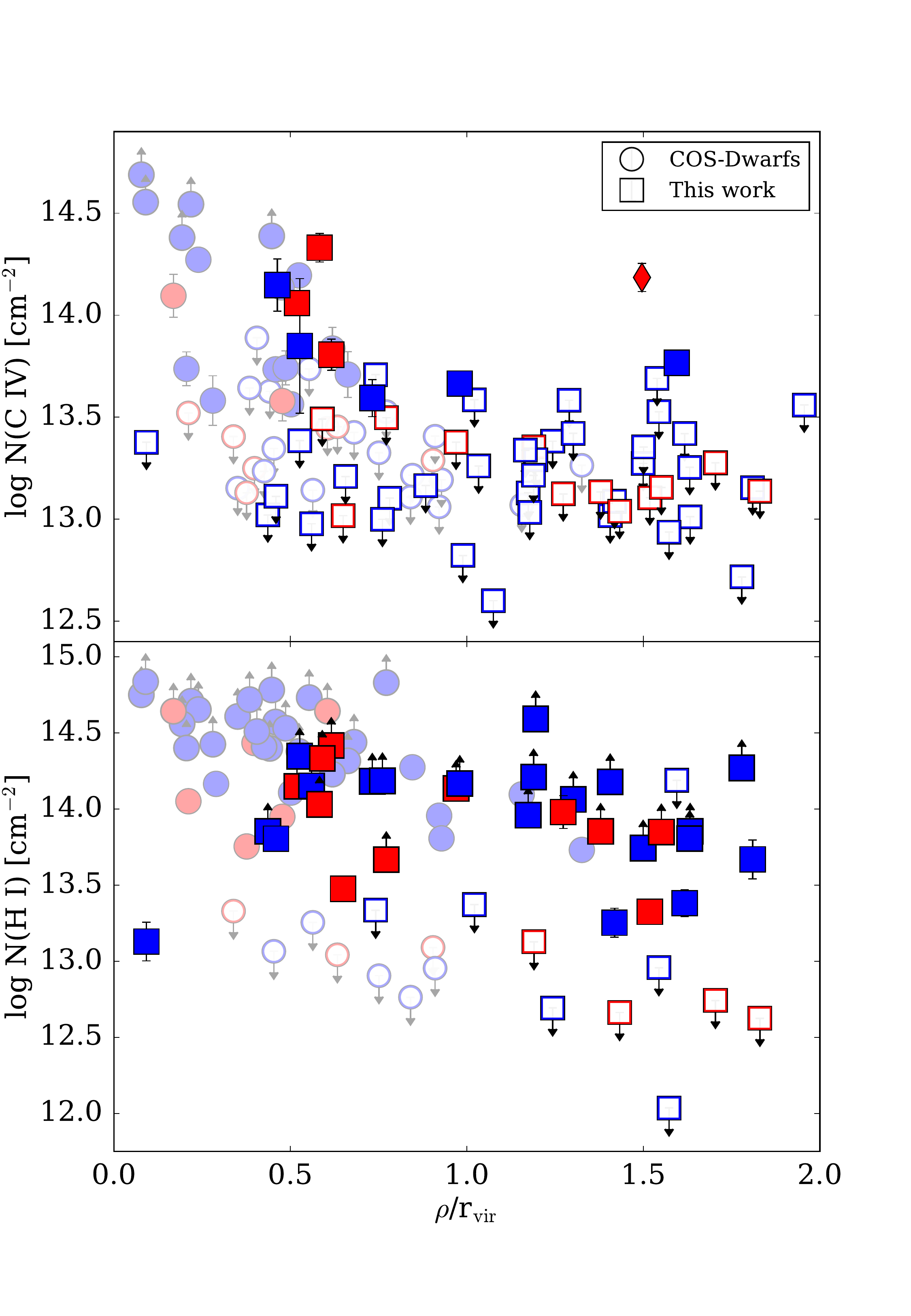}
\caption{The \cfour\ (top) and \hone\ (bottom) column densities of our low-z absorber sample (squares) as functions of impact parameter of the associated galaxies selected by fraction of the galaxy virial radius using a galaxy magnitude limit of $\mathcal{M}_r \leq -16.4$. The COS-Dwarfs sample \citep[fainter circles; ][]{Bordoloi:2014lr} is also plotted for comparison. The galaxy/absorber pair indicated with a diamond in the top panel has the greatest velocity separation of all pairs plotted, with $\delta v >600$ km/s.  Red and blue symbols denote passive and star-forming galaxies, respectively. The open symbols with downward arrows correspond to  3-$\sigma$ upper limits on the absorption; filled symbols with upward arrows indicate lower limits measured using the apparent optical depth on saturated lines. }
\label{fig:rvirimpact}
\end{figure}

Using the selection procedures described above, we produced samples of galaxy/absorber pairs imposing a galaxy luminosity cut of $\mathcal{M}_r \leq -16.4$, corresponding to the completeness limit of SDSS (see Section \ref{sec:galsurveys}). Figures \ref{fig:kpcimpact} and \ref{fig:rvirimpact} show the absorber \cfour\ and \hone\ column densities (including 3-$\sigma$ upper limits) as a function of impact parameter; Figure \ref{fig:kpcimpact} presents the proper-distance selected sample, and Figure \ref{fig:rvirimpact} shows the virial-radius selected sample.  In these figures, we differentiate between passive and star-forming galaxies using the sSFR separation described in Section \ref{sec:SFR}.  For the detected absorbers in our sample, the column densities shown represent the total column densities summed over all velocity components in each absorber. The \cfour~column density upper limits shown were measured by integrating the \cfour~\lam~1548 apparent column density profiles \citep{Savage:1991vn} within $\pm 50$ km/s at the redshifts where no \cfour~absorber was detected.   The largest impact parameters in these plots correspond to the absorber at $z_{abs} = 0.00261$ in the sightine of QSO J0925+4004, for which no galaxy was found closer than $|\delta v|$ = 700 km/s. For comparison, we show in the top panels of Figs. \ref{fig:kpcimpact} and \ref{fig:rvirimpact} the \cfour~absorber-galaxy pairs from the COS-Dwarfs survey \citep[faint circles;][]{Bordoloi:2014lr} alongside our blindly selected dataset.  

For the proper distance-selected sample from our blind survey, we detect \cfour\ absorption within 160 kpc of 6 out of 19 galaxies ($32^{+11}_{-10}$\%).  Within the measurement uncertainties, this detection rate agrees with the COS-Dwarfs result of 17 detections out of 43 galaxies targeted ($39^{+8}_{-7}$\%).   For our virial radius-selected sample, we obtain 7 detections around 21 galaxies ($33^{+11}_{-9}$\%) probed within 1 virial radius; COS-Dwarfs yielded 17 detections within 1 virial radius of 41 galaxies targeted ($41^{+8}_{-7}$\%).  Our blind survey also extends the CGM probed by \cfour~absorption out to impact parameters of 400 kpc and beyond, and we find absorbers out to $\sim350$ kpc  and $>1.5~r_{\rm vir}$ from the nearest galaxy.  However, beyond 160 kpc and 1 \rvir, the detection rate dramatically decreases for the proper distance- and virial radius-selected samples, respectively:  two detections out of a possible 19 ($10^{+10}_{-5}$\%) occur in the 160-250 kpc region;  we report 0/16 detections ($0^{+6}_{-0}$\%) at 1-1.5 \rvir\ and 1/12 ($8^{+11}_{-5}$\%) at 1.5-2 \rvir.  Covering fractions under certain selection criteria for both the proper distance selection and the virial radius selection are shown in the top two sections of Table 3.  

The detection at the largest impact parameter in the proper distance-selected sample (368 kpc) occurs for the absorber at $z_{abs}=0.00261$ in the J0925+4004 sightline, and this absorber is associated with the same galaxy, NGC 2844, in both selection methods (but note the caveat in Section \ref{sec:selection} regarding this association).  NGC 2844 also has a very large velocity separation with the absorber (-701 km/s) but serves as the closest in impact parameter with the smallest velocity offset.

Two notable points are immediately apparent.  First, as seen in Figure \ref{fig:kpcimpact}, the detections of \cfour~absorption are predominantly associated with galaxies, most often within 200 kpc of a galaxy.  This result was in essence first demonstrated for strong absorbers ($W_{\rm 1548} \gtrsim 200$~m\AA) by \citet{Chen:2001lr} using data from the \textit{Faint Object Spectrograph} on HST \citep{Bahcall:1993zr}. Our survey enables sensitivity to much weaker absorbers ($W_{\rm 1548} \lesssim 100$~m\AA) and fainter galaxies, and we indeed find that this close association persists.  However, the nearest luminous galaxies to the absorbers, such as those selected by virial radius, have much larger impact parameters for 3/9 absorbers ($\rho > 250$ kpc; Tables 1 and 2).   

Second, some metal-enriched gas may arise well beyond the virial radius of a galaxy.  Absorbers at large distances have been previously reported \citep{Stocke:2013mz, Tripp:2006qy, Johnson:2013lr}; however, the COS-Dwarfs sample showed no \cfour~detections beyond approximately $0.5~r_{vir}$ \citep{Bordoloi:2014lr}, corresponding to $0.66~r_{vir}$ using the cosmology and velocity corrected distances adopted here and reflected in Figs. \ref{fig:kpcimpact} and \ref{fig:rvirimpact}.  We further compare these results with previous studies in Section \ref{sec:comparison}.

\subsubsection{\hone\ column density}
\label{sec:HIcolprof}
The bottom panels of Figs. \ref{fig:kpcimpact} and \ref{fig:rvirimpact} show the corresponding \hone\ column density measurements for the same galaxies appearing in the \cfour\ profiles in the top panels.  The \hone\ identifications and measurements suffer from two complications that do not arise for \cfour: (1) At the redshifts we are probing in this analysis, only the \lya\ line, which is typically saturated and only yields a lower limit on N(\hone), falls in the COS bandpass.  (2) The \lya\ lines at the redshifts of these $z<0.015$ galaxies often fall within the Galactic \lya\ profile where the Milky Way damping wings reduce the S/N or the flux disappears altogether.  \hone\ detections are shown in Figs. \ref{fig:kpcimpact} and \ref{fig:rvirimpact} where \lya\ absorption was identified within 600 km/s of a galaxy's redshift. We attempted to corroborate \lya\ detections with other metal lines where possible to ensure that the line in question was not associated with another known system in the spectrum.  If the flux values in the core of line profile were greater than the corresponding errors, we fitted the line with a Voigt profile to obtain a column density.  Otherwise, a lower limit on the column density was measured using the apparent optical depth method (AODM) \citep{Savage:1991vn}.  For nondetections, we measured upper limits using the AODM with the error vector over a velocity range of $\pm$ 50 km/s centered on the rest frame of the detected galaxy. \hone\ data for the COS-Dwarfs galaxy/absorber pairs are adopted from Bordoloi et al. (2016, in preparation). 

Figures \ref{fig:kpcimpact} and \ref{fig:rvirimpact} show a markedly higher detection rate for \lya than for \cfour, at all impact parameters.  Higher detection rates of \hone\ over metal ions have been reported in many absorber/galaxy studies \citep[e.g.,][]{Wakker:2009fr,Prochaska:2011yq,Stocke:2013mz}, and we briefly contrast the emerging views of the CGM traced by \cfour\ and \hone.  

First, we underscore that while the CGM is quite patchy to \cfour, \hone\ is nearly ubiquitous.  If nearly all galaxies have gaseous halos, as indicated by the very high covering fraction within 1 $r_{\rm vir}$ ($0.94^{+0.04}_{-0.10}$ for N(\hone)~$>10^{13.5} ~\cmt$), the gas traced by \cfour\ may therefore possess a `special' set of ionizing conditions sufficient for a detectable fraction of the carbon to be triply ionized.  Another possibility is that the clouds traced by \cfour\ represent a more metal-enriched sample of the larger population of halo clouds; however, as also pointed out by \citet{Stocke:2013mz}, the \cfour\ clouds may simply comprise clouds of similar metallicity as the \hone-only detections but are simply more massive.  Lastly, many of the \hone-only detections might be metal-poor intergalactic gas, while the \cfour\ absorbers tend to trace metal-enriched circumgalactic gas.

Second, among the 26 \hone\ clouds we detect within 2 $r_{\rm vir}$ of a galaxy, only 5 have well constrained \hone\ column densities that are less than log N(\hone)$~= 10^{13.5}~\cmt$. \citet{Chen:2005kx} conducted a cross-correlation analysis between \hone\ absorbers and galaxies at $z<0.5$ and report that N(\hone)$~> 10^{14}~\cmt$ absorbers have a significant cross-correlation signal with emission line galaxies that rivals the autocorrelation of galaxies with themselves.  However, they find that N(\hone)$<  10^{13.6}~\cmt$ absorbers have a random distribution with galaxies relative to galaxies with themselves.  Therefore, the weaker \hone\ absorbers that we detect are similar (at least in \hone\ column density) to those that statistically bear the mark of run-of-the-mill IGM clouds.   Figure \ref{fig:kpcimpact} shows several \hone\ detections at $\rho \gtrsim 300$~kpc but only 1 \cfour\ detection at $\rho \gtrsim 300$~kpc.  Half of these \hone\ detections have N(\hone)$ \lesssim 10^{13.7} \cmt$, which coupled with the decreased \cfour\ detection rate suggests that the conditions giving rise to the characteristic differences between the CGM and IGM have generally transitioned to IGM by $\rho \sim 300$~kpc. 

\begin{figure}[t!]
\centering
\includegraphics[width=.52\textwidth]{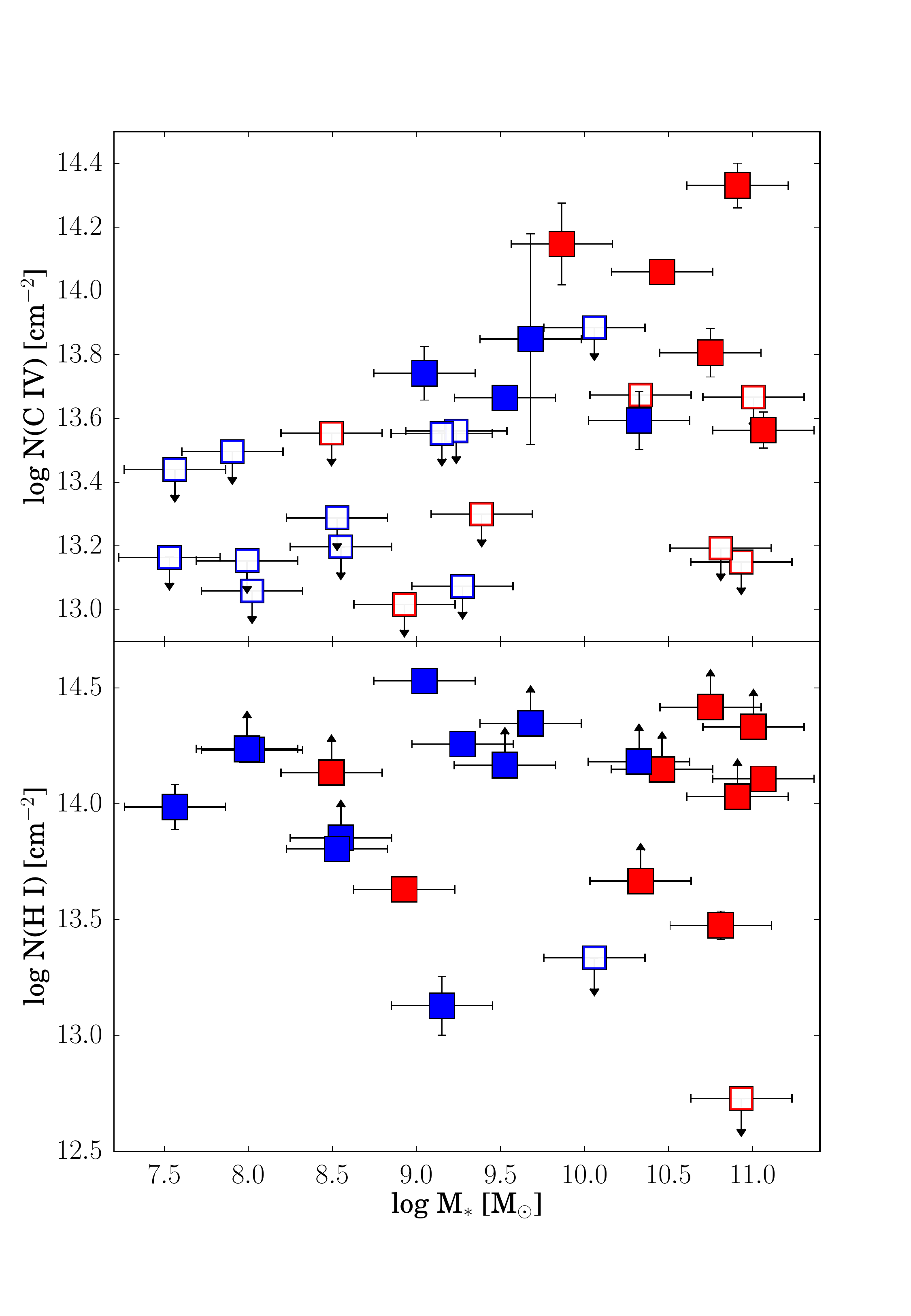}
\caption{The \cfour~ (top) and \hone~ (bottom) column density as a function of stellar mass of the highest-mass galaxy within 1 virial radius.  Red and blue symbols correspond to passive and star-forming galaxies, respectively.  The downward arrows correspond to  3-$\sigma$ upper limits on the absorption. We note a dramatically increased detection rate of \cfour\ within 1 \rvir\ of galaxies with log $M* > 9.5 ~M_{\odot}$.  The open symbols with downward arrows correspond to  3-$\sigma$ upper limits on the absorption; filled symbols with upward arrows indicate lower limits measured using the apparent optical depth on saturated lines.}
\label{fig:masscol_rvir}
\end{figure}

\subsubsection{The CGM as a function of stellar mass}
\label{sec:massciv}

Our blind survey provides a galaxy sample independent of any mass selection criteria and that spans a wide dynamic range in mass.  We now investigate whether the CGM absorption, in terms of \cfour\ and \hone, depends on galaxy stellar mass.  For this analysis, we selected galaxies in a similar manner to the virial-radius selection described in Section \ref{sec:selection} but with one key difference. When multiple galaxies were probed within 1 \rvir\ by a QSO sightline and those galaxies had redshifts within $\sim 600$ km/s of one another, we selected the most massive of those galaxies.   As shown in Figure \ref{fig:absmaps}, low-mass galaxies probed within 1 \rvir\ are frequently satellite galaxies  within 1 \rvir\ of a more massive galaxy.  By choosing the most massive galaxies within 1 \rvir, we ensure a fair comparison between the halos of low-mass and higher-mass galaxies because the halo properties of these low-mass galaxies are less likely to be dominated by a massive counterpart, i.e., the low-mass galaxies should be centrals of their halos.  Because their targeting procedure does not introduce bias into this analysis, we include the COS-Dwarfs data here; however, we subjected those galaxies/absorbers to the same procedure as the rest of the data.  In certain cases, a more massive galaxy than that targeted by COS-Dwarfs was present at a similar redshift, and we included the more massive galaxy in our sample.  We note that the results of this section are qualitatively insensitive to small changes in the 1 \rvir\ selection criterion, although increasing the selection beyond 1 \rvir\ produces more non-detections in the both the low- and high-mass regimes (consistent with the previous section's results). As in the previous section, this analysis is limited to $z<0.015$ for survey completeness.

Figure \ref{fig:masscol_rvir} shows the \cfour\ and \hone\ column densities as a functions of galaxy stellar mass according to the above selection procedure.  We note that the detection rate of \cfour\ sharply increases for $M_* >10^{9.5} ~M_{\odot}$ galaxies.  Using a detection threshold of log $\cfourcoleq > 13.5~\cmt$, we detect \cfour~within 1 virial radius of 8 out of 10 galaxies ($80^{+10}_{-15}$\%) and 1 out of 11 galaxies ($9^{+12}_{-5}$\%) in the $M_* >10^{9.5} ~M_{\odot}$ and $M_* <10^{9.5} ~M_{\odot}$ bins, respectively.  This result is qualitatively consistent with the \osix~dependence observed by \citet{Prochaska:2011yq}, although our mass bins are substantially less populated.  To test the significance of this difference, we employ a Fisher's exact test, which is well suited to small sample sizes, assuming that the variables are categorical with galaxy mass categories of $> 10^{9.5}$ and $<10^{9.5} ~M_{\odot}$ and categories for the detection and nondetection of \cfour. As a result, we reject the null hypothesis that the difference in the fraction of detections between galaxy masses is purely random with confidence $>99.7\%$.  We also estimated the significance of this result by taking the same distribution of galaxy masses as shown in Figure \ref{fig:masscol_rvir}, randomly assigning \cfour\ detection or nondetection status to each each galaxy based on the covering fraction of $\sim 40\%$ reported by \citet{Bordoloi:2014lr}, and producing $10^6$ Monte Carlo resamples. Based on the probability of reproducing the observed distribution of \cfour\ detections and nondetections, we reject the null hypothesis that the rates observed between the $> 10^{9.5}$ and $<10^{9.5} ~M_{\odot}$ galaxies occur at random with $>99.9\%$ confidence.

For comparison, the bottom panel of Figure \ref{fig:masscol_rvir} shows the same mass-column density relationship for neutral hydrogen within 1 \rvir\ of the same $z<0.015$ galaxies.  The issues with \hone\ measurements discussed in Section \ref{sec:HIcolprof} also arise here and result in a smaller sample being represented in the bottom panel than the top panel (for \cfour).  However, \hone\ \emph{is} detected even in the halos where we do not detect \cfour.  Assuming a detection threshold of log $\cfourcoleq > 13.5~\cmt$, the $<10^{9.5} ~M_{\odot}$ bin has an \hone\ detection rate within 1 \rvir\ of  $100^{+0}_{-10}$\%, not statistically different from that for more massive galaxies, $82^{+9}_{-14}$\% of which yield \hone\ detections.  As for the previous sections' analysis, these covering fractions are summarized in Table 3.  Nevertheless, \lya\ is a much stronger transition than the \cfour\ lines and is predominately saturated in our data; thus, an \hone-mass dependence may still be present but could be obscured by our inability to precisely constrain the column densities in the strong \hone\ lines.  

A number of scenarios may be invoked to explain the difference in \cfour~detection between the galaxy mass bins.  These include, but are certainly not limited to, the following: 1) The gas in the CGM of dwarfs is sufficiently less dense or less self-shielded as to allow more energetic photons to ionize the \cfour\ to a higher ionization state(s). In this hypothesis, the \hone\ would be more ionized as well, but \hone\ remains detectable even when the \hone\ ionization fraction is quite small, so the \cfour\ absorption could disappear while the \hone\ persists. Better constrained \hone\ column densities may provide insight here as whether a significant difference exists in the \hone\ column densities of Figure \ref{fig:masscol_rvir} between the two mass bins. 2) Because of their shallower potential wells, the lowest mass dwarf galaxies are better able to expel their metal-enriched gas into IGM  regions where the gas is further ionized and/or falls below the column  density detection threshold. 3) The gas in a lower ionization stage due to a lower virial temperature of the low-mass halo, and the carbon is better traced by \cthree\ (not covered in our data) than \cfour.   4) Less massive galaxies have less mass in their CGM and shorter sightline paths through their halos; therefore, given similar underlying physical conditions and feedback behavior as in more massive galaxies, the column density (and detectability) of any species will be lower.  This would affect both \cfour\ and \hone, but because the \hone\ lines are predominately saturated, it could be difficult to see this effect in the \hone\ data.

\subsection{Galaxy environments of \cfour~absorbers}
\label{sec:envirociv}
As shown in our work and in the previous studies discussed, \cfour~absorbers are typically found to be coincident with nearby galaxies \citep[with a few exceptions, e.g., ][]{Tripp:2006qy}.  As shown in Figure \ref{fig:absmaps}, the \cfour~absorbers clearly occupy a variety of environments, from one or two nearby galaxies to relatively well-populated groups, and many absorbers occur within the virial radii of multiple galaxies.  Our small galaxy/absorber subsample might represent a wide diversity in the physical nature of the gas detected. While environmental effects are conspicuous in galaxy disks and central regions, the mechanisms involved must also be felt in the intermediary CGM and, therefore, impact the feeding and outflow processes occurring there.  For instance, if galaxies reside in a larger group halo, would-be cold-mode infalling gas could be shock heated even when the individual galaxies' subhalo is below the threshold mass ($M_{halo} = 10^{11.4} ~M_{\odot}$) at which an individual halo would dynamically shock-heat accreting intergalactic gas.  Furthermore, tidal forces may assist the mass transfer of metal-enriched outflows to leave the CGM.  Thus, we now leverage our low-z absorber sample, for which we have rich galaxy survey data from SDSS, to investigate the role of environment on the gas traced by \cfour.  

The analyses herein leverage the full NASA-Sloan atlas and all \cfour\ absorbers at $z<0.055$ within our QSO sightline sample.  As in Section \ref{sec:massciv}, the COS-Dwarfs sample introduces no inherent bias to the current environmental analysis, and these galaxies/absorbers are included.

\begin{figure}[t!]
\centering
\includegraphics[width=.52\textwidth]{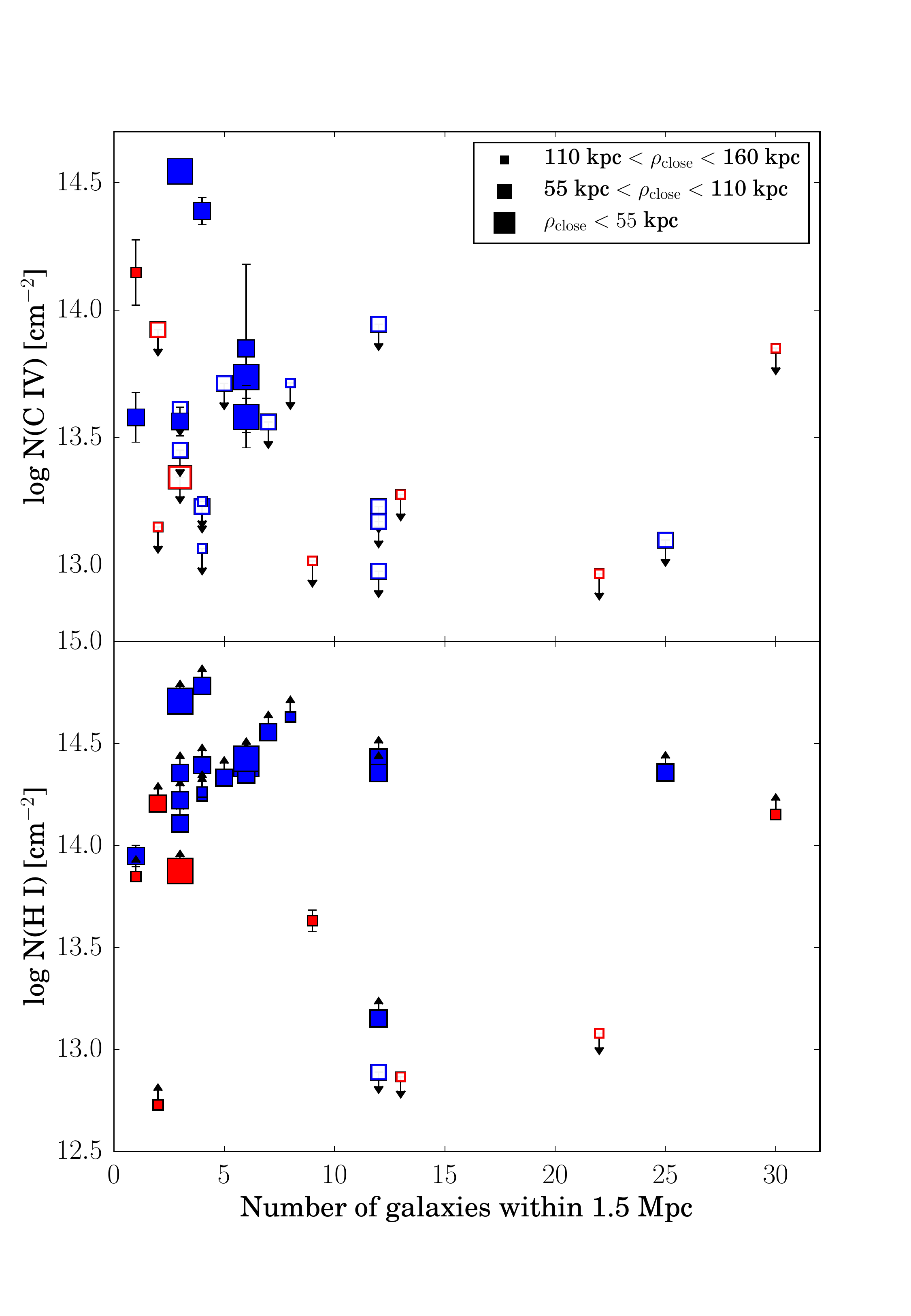}
\caption{(Top) \cfour~column density as a function of the number of galaxies within 1.5 Mpc of the QSO sightline, where filled squares represent detections of \cfour~and open squares represent nondetections; red and blue squares correspond to the passive and star-forming galaxies, respectively. Shown are the systems where at least one $\mathcal{M}_r < -19$ ($M_*\gtrsim10^{9.5}~\msuneq$) galaxy falls within an impact parameter $\rho<160$ kpc. The x-axis represents the number of galaxies (once again with $\mathcal{M}_r < -19$) within a 1.5 Mpc projected aperture centered on the sightline and $\Delta z = 0.0033$ of the nearest galaxy to the sightline. Upper limits shown on the column density are 3-$\sigma$.  Note the lack of \cfour~detections where the number of galaxies within 1.5 Mpc exceeds 7. Symbol sizes are scaled according to impact parameter (larger symbols have smaller impact parameters) of the closest galaxy ($\leq 160$ kpc) to show that impact parameter effects alone do not account for the detection or nondetection of \cfour. (Bottom) \hone\ column density as a function of the same environmental metric. Unlike \cfour, \hone\ is detected at even the highest densities probed, although the only \hone\ nondetections do at occur at relatively high density.}
\label{fig:gals1000kpc}
\end{figure}

\subsubsection{Fixed-aperture Galaxy Density}

To investigate how the detection of \cfour~depends on the local galaxy environment, we first employ a straightforward, sightline-centric fixed-aperture number density, i.e., how does the observed \cfour~column density (or upper limit for nondetections) depend on the number of nearby galaxies within some projected distance and velocity tolerance?  \citet{Muldrew:2012qy} employed a large volume of simulated dark matter halos to compare several metrics commonly used to quantify environment.  In general, fixed-aperture methods and nearest-neighbor distances are found to probe different scales of environment (individual halos to large-scale structure), and these dependencies are sensitive to the parameters chosen.  Both \citet{Muldrew:2012qy} and \citet{Haas:2012kq} find that fixed-aperture densities correlate well with halo mass on scales of rich groups to clusters using apertures of $\sim1 h^{-1}$ Mpc and $\delta v = \pm1000$ km/s.  However, smaller halos ($\mhaloeq < 10^{13} \msuneq$) are not well differentiated by this metric.

As stated in Section \ref{sec:abssample}, we use galaxies with k-corrected absolute r-band magnitudes of $\mathcal{M}_r \leq -19$ as tracers of the local density.  As shown in Section \ref{sec:massciv}, \cfour\ absorption undergoes an increase in detection within the virial radius at $\mstareq > 10^{9-9.5} \msuneq$.  The distribution of galaxies at $\mstareq \sim 10^{9.5}$ peaks at $\mathcal{M}_r \sim -19$, which is conveniently 0.1 mag brighter than the expected spectroscopic completeness limit at $z \sim 0.055$, the upper redshift limit of the NASA-Sloan Atlas.   Therefore, we select from our entire dataset the sightlines and redshifts where galaxies more luminous than $\mathcal{M}_r < -19$ occur within impact parameters of $\rho \leq 160$ kpc.  This distance is approximately the virial radius of a $\mstareq = 10^{9.5} \msuneq$ galaxy; furthermore,  \cfour~detection declines at larger impact parameters \citep{Bordoloi:2014lr, Liang:2014kx}.    Finally, we counted the number of $\mathcal{M}_r < -19$ galaxies,  N$_{1500}$, within $\rho < 1500$ kpc (approx. 1 $h^{-1}$ Mpc in the adopted cosmology) and $|\delta z| < 0.0033$ (approx. 1000 km/s) in accordance with the results of \citet{Muldrew:2012qy} and \citet{Haas:2012kq}.  Our results do not qualitatively change under  small deviations ($\sim 10\%$) of these parameters.  Maps of the galaxy environments included in this analysis are provided in the Appendix; also included are plots showing overdensities at various redshifts along the QSO sightline.   

Figure \ref{fig:gals1000kpc} shows detections and non-detections of \cfour~at redshifts $z<0.055$ where at least one $\mstareq \gtrsim 10^{9.5} \msuneq$ galaxy has $\rho \leq 160$ kpc.   A density threshold is apparent, as the regions of highest density with N$_{1500}\gtrsim7$ do not contain \textit{any} \cfour\ absorbers above the detection limits imposed by the COS spectra.  In contrast with \cfour, but as observed with the mass dependence discussed in Section \ref{sec:massciv}, the bottom panel of Figure \ref{fig:gals1000kpc} shows that \hone\ is detected at the highest densities where \cfour\ is not.  However, the only 3 nondetections of \hone\ do occur at N$_{1500} > 10$, which may hint at a density dependence, but larger samples with more precise \hone\ column densities are required to test this trend.

To quantify the statistical significance of this result, we compared the \cfourcol\ measurements for galaxies in environments above and below the apparent N$_{1500}=7$ density threshold seen in Figure \ref{fig:gals1000kpc}.  The log-rank nonparametric test, which accepts censored data, with 3-sigma upper limits on the column density for nondetections, indicates that we may reject the null hypothesis that the two samples are drawn from the same distribution with $>$98\% confidence.  Alternatively, we compared the two detection rates adopting a detection threshold of log N(\cfour) = 13.5 cm$^{-2}$ using the Newcombe-Wilson method for the difference between two proportions; the test yields a 99\% confidence that we may reject the null hypothesis that the detection rate is equal above and below the threshold value.   For reference, Table 3 includes the covering fractions for \cfour\ and \hone\ in the two N$_{1500}$ density bins. 

We qualify that we have assessed the statistical significance adopting an apparent threshold value of N$_{1500}$. The apparent threshold value depends on, e.g., increasing or decreasing the outer impact parameter from 1500 kpc. Smaller choices of outer impact parameter decrease the apparent threshold value by one or two galaxies, bringing more nondetections to lower densities.  However, a threshold density remains apparent unless the chosen outer impact parameter is much smaller than 1000 kpc, at which scales the fixed-aperture density may not be probing the full extent of more massive halos.  Conversely, larger choices of outer impact parameter aperture probe scales beyond individual halos, and the detections at low density in Figure \ref{fig:gals1000kpc} would move to higher density because of contamination from other halos in the density metric.

Figures \ref{fig:kpcimpact} and \ref{fig:rvirimpact} show that the metal-enriched gas traced by \cfour\ absorption preferentially resides within projected distances of $\sim 200$ kpc from nearby galaxies.  However, the dearth of \cfour\ detections in dense environments as shown in Figure \ref{fig:gals1000kpc} suggests that the presence of many galaxies near a sightline does not have an additive effect on the presence of \cfour-traced gas, as these environments should provide more potential sources for the metals.   Therefore, the absorption line data may be reflecting changing physical conditions of the CGM influenced by the larger scale environment.  We return to this assertion for further discussion in Section \ref{sec:groupclusterdiscuss}. 

\subsubsection{Group Dark Matter Halo Mass}

While Figure \ref{fig:gals1000kpc} suggests a \cfour~detection threshold in terms of the local galaxy density quantified by simply counting galaxies, the question remains whether the galaxies whose CGM is probed falls within the same larger dark matter halo as the surrounding galaxies counted.  Therefore, we have cross-matched the galaxies initially selected in our fixed-aperture analysis, i.e., those with $\rho<160$ kpc, with the group catalog of \citet{Yang:2007kl} (using their version updated for SDSS DR7) to obtain the halo masses (\mhalo) of the groups in which these galaxies reside.  We assigned halo masses as follows: (1) For the galaxies that were directly classified as group members by \citet{Yang:2007kl}, we assigned group halo masses from the catalog.  (2) If the galaxies were not identified as group members and were not projected within $1~\rvireq$ of any group in the catalog, we assigned the halo mass from abundance matching as described in \ref{sec:massesrvir}.  (3) Galaxies that were not identified as group members but were at projected distances within $1~\rvireq$ of a group were assigned the mass of that group.  

\begin{figure}[t!]
\centering
\includegraphics[width=.42\paperwidth]{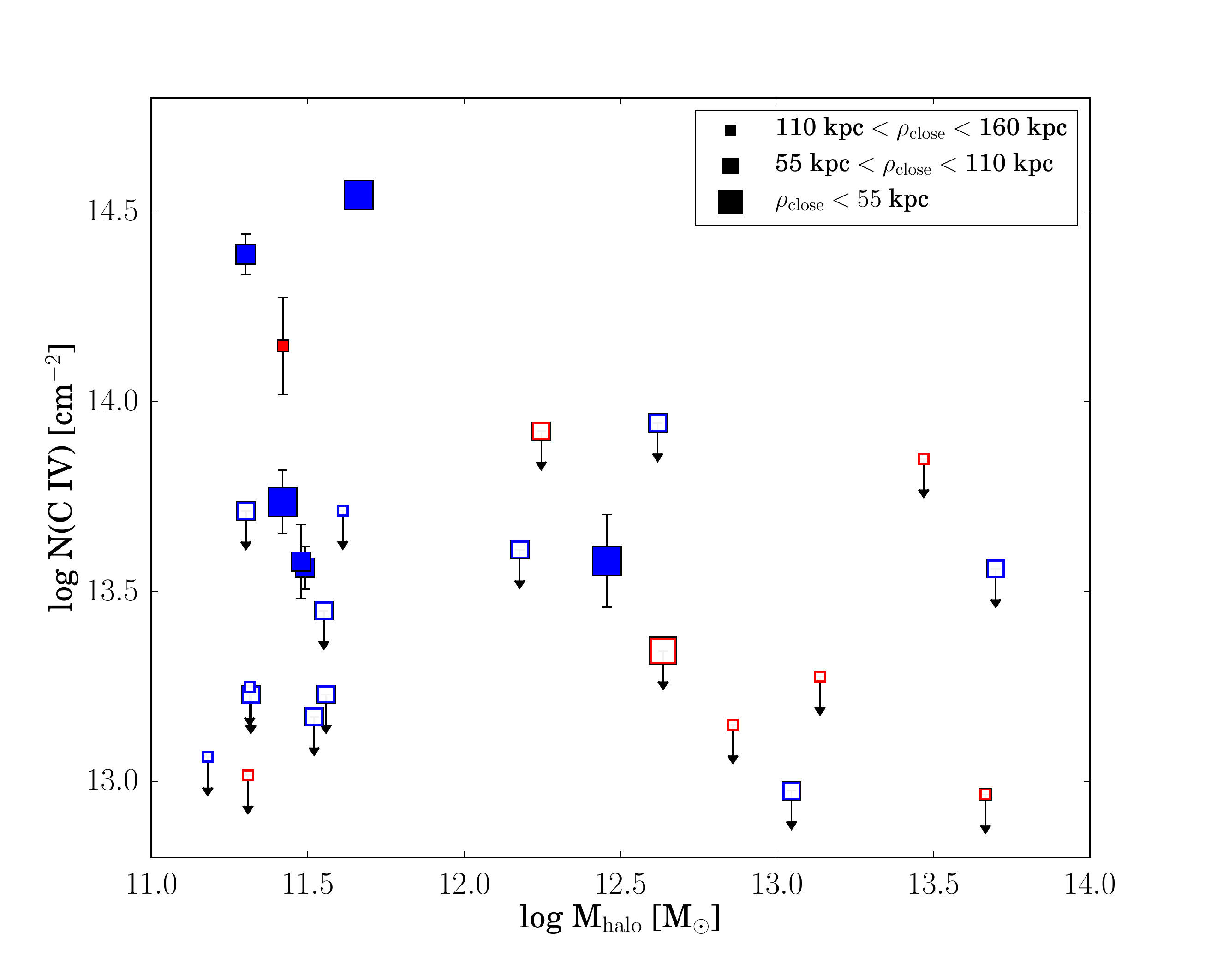}
\caption{The \cfour\ column density as a function of dark matter group halo mass for the CGM absorbers and nondetections plotted in Fig. \ref{fig:gals1000kpc}. The halo masses were adopted from the \citet{Yang:2007kl} group catalog for the galaxies within 160 kpc of the QSO sightlines or from halo abundance matching for galaxies not in groups.  As in Figure \ref{fig:gals1000kpc}, the larger symbols indicate smaller impact parameters to the closest galaxy. The open symbols with downward arrows correspond to  3-$\sigma$ upper limits on the absorption; filled symbols with upward arrows indicate lower limits measured using the apparent optical depth on saturated lines.  Consistent with the galaxy counts plotted in Fig. \ref{fig:gals1000kpc}, the galaxies residing in halos of M$_h \gtrsim 10^{12.7} M_\odot$ do not show \cfour\ absorption in their CGM.  }
\label{fig:c4halomass}
\end{figure}

The resulting \cfourcol-\mhalo\ relation is shown in Figure \ref{fig:c4halomass}; as for the most populated local regions depicted in Figure \ref{fig:gals1000kpc}, we do not detect \cfour\ in the CGM of galaxies residing in the largest-\mhalo\ groups (\mhalo$\sim 10^{12.7}$ \msun).   The group catalog only includes galaxies at $z \geq 0.01$, so any non-isolated galaxies at $z<0.01$ shown in Figure \ref{fig:gals1000kpc} are omitted from this figure.  The only $z<0.01$ galaxy included here contains no other galaxies within the 1500-kpc aperture, and we have assigned its halo mass through abundance matching as with the other isolated galaxies.   

The gas traced by \cfour~in the low-z, inner CGM is likely subject to a number of influences, evidenced by	 (1) trends of decreasing column densities for several ions with increasing impact parameter and increasing column density ratios of higher to lower ions with increasing impact parameter \citep{Bordoloi:2014lr,Liang:2014kx} and (2) the galaxy mass dependence reported in Section \ref{sec:massciv}.  The results presented in this section suggest that a third environmental influence must be added as well.  We have attempted to control for impact parameter- and mass-dependent effects to some degree by requiring that at least one $\mathcal{M}_r \leq -19$ galaxy fall within $\rho = 160$ kpc of the QSO sightline, but much larger galaxy/absorber samples are required to separate these effects while simultaneously controlling for galaxy density, mass, etc.

\begin{deluxetable*}{ccccccc} 
\centering 
\tabletypesize{\scriptsize}  
\tablewidth{0pt}  
\tablecaption{Covering fractions of \cfour\ and \hone\ absorption with respect to impact parameter, galaxy mass, and environment.} 
\tablehead{  Ion &  Detection Threshold &   Redshift range &  Selection Criteria &  N$_{det}$ &  N$_{tot}$ &  $f_c$ \\ \ & (log N(X) [$\cmt$]) & \ & \ & \ & \ & (\%) } 
\hline 
\startdata  
\multicolumn{7}{c}{Impact parameter distributions for galaxies with $\mathcal{M}_r \leq -16.4$ (proper distance-selected)} \\ 
\hline 
\cfour & 13.5 & $z \leq 0.015$ & $\rho < 160~{\rm kpc}$ & 6 & 19 & 32$^{+11}_{-10}$ \\
\cfour & 13.5 & $z \leq 0.015$ & $\rho < 100~{\rm kpc}$ & 3 & 8 & 38$^{+18}_{-15}$ \\
\cfour & 13.5 & $z \leq 0.015$ & $100~{\rm kpc} \leq \rho < 200~{\rm kpc}$ & 5 & 21 & 24$^{+10}_{-8}$ \\
\cfour & 13.5 & $z \leq 0.015$ & $200~{\rm kpc} \leq \rho < 300~{\rm kpc}$ & 0 & 16 & 0$^{+6}_{--0}$ \\
\cfour & 13.5 & $z \leq 0.015$ & $300~{\rm kpc} \leq \rho < 400~{\rm kpc}$ & 0 & 12 & 0$^{+8}_{--0}$ \\
\hone & 13.0 & $z \leq 0.015$ & $\rho < 160~{\rm kpc}$ & 12 & 13 & 92$^{+5}_{-11}$ \\
\hone & 13.0 & $z \leq 0.015$ & $\rho < 100~{\rm kpc}$ & 7 & 7 & 100$^{+0}_{-13}$ \\
\hone & 13.0 & $z \leq 0.015$ & $100~{\rm kpc} \leq \rho < 200~{\rm kpc}$ & 10 & 13 & 77$^{+10}_{-13}$ \\
\hone & 13.0 & $z \leq 0.015$ & $200~{\rm kpc} \leq \rho < 300~{\rm kpc}$ & 13 & 17 & 76$^{+9}_{-12}$ \\
\hone & 13.0 & $z \leq 0.015$ & $300~{\rm kpc} \leq \rho < 400~{\rm kpc}$ & 5 & 8 & 62$^{+15}_{-18}$ \\
\hline 
\multicolumn{7}{c}{Impact parameter distributions for galaxies with $\mathcal{M}_r \leq -16.4$ (virial radius-selected)} \\ 
\hline 
\cfour & 13.5 & $z \leq 0.015$ & $\rho < 1~\rvireq $ & 7 & 21 & 33$^{+11}_{-9}$ \\
\cfour & 13.5 & $z \leq 0.015$ & $1~\rvireq \leq \rho < 2~\rvireq $ & 1 & 28 & 4$^{+5}_{-2}$ \\
\hone & 13.0 & $z \leq 0.015$ & $\rho < 1~\rvireq $ & 13 & 14 & 93$^{+4}_{-10}$ \\
\hone & 13.0 & $z \leq 0.015$ & $1~\rvireq \leq \rho < 2~\rvireq $ & 13 & 22 & 59$^{+10}_{-11}$ \\
\hline 
\multicolumn{7}{c}{Dependence on galaxy mass within $\rho = 1~\rvireq$ } \\ 
\hline 
\cfour & 13.5 & $z \leq 0.015$ & $\mstareq < 9.5~\msuneq$ & 1 & 11 & 9$^{+12}_{-6}$ \\
\cfour & 13.5 & $z \leq 0.015$ & $\mstareq \geq 9.5~\msuneq$ & 8 & 10 & 80$^{+10}_{-15}$ \\
\hone & 13.0 & $z \leq 0.015$ & $\mstareq < 9.5~\msuneq$ & 9 & 9 & 100$^{+0}_{-10}$ \\
\hone & 13.0 & $z \leq 0.015$ & $\mstareq \geq 9.5~\msuneq$ & 9 & 11 & 82$^{+9}_{-14}$ \\
\hline 
\multicolumn{7}{c}{Dependence on galaxy environment given an $\mathcal{M}_r \leq -19$ galaxy within $\rho =  160$ kpc} \\ 
\hline 
\cfour & 13.5 & $z \leq 0.055$ & $\mathcal{N}_{1500} \leq 7$ & 8 & 14 & 57$^{+12}_{-13}$ \\
\cfour & 13.5 & $z \leq 0.055$ & $\mathcal{N}_{1500} > 7$ & 0 & 7 & 0$^{+13}_{-0}$ \\
\hone & 13.0 & $z \leq 0.055$ & $\mathcal{N}_{1500} \leq 7$ & 16 & 16 & 100$^{+0}_{-6}$ \\
\hone & 13.0 & $z \leq 0.055$ & $\mathcal{N}_{1500} > 7$ & 7 & 10 & 70$^{+12}_{-16}$ \\
\hline 
\multicolumn{7}{c}{Dependence on group dark halo mass given an $\mathcal{M}_r \leq -19$ galaxy within $\rho =  160$ kpc} \\ 
\hline 
\cfour & 13.5 & $z \leq 0.055$ & $\mhaloeq < 12.5 \msuneq$ & 7 & 14 & 50$^{+13}_{-13}$ \\
\cfour & 13.5 & $z \leq 0.055$ & $\mhaloeq \geq 12.5 \msuneq$ & 0 & 5 & 0$^{+17}_{-0}$ \\
\hone & 13.0 & $z \leq 0.055$ & $\mhaloeq < 12.5 \msuneq$ & 16 & 17 & 94$^{+4}_{-9}$ \\
\hone & 13.0 & $z \leq 0.055$ & $\mhaloeq \geq 12.5 \msuneq$ & 5 & 6 & 83$^{+10}_{-20}$ 
\enddata 
\label{tab:covfracs} 
\end{deluxetable*}

\section{Discussion}
\label{sec:discussion}
\subsection{Faint dwarf galaxies as underrepresented sources of \cfour\ absorbers}
As shown in Figure \ref{fig:compareselections}, for 5/9 absorbers, the galaxies with the smallest impact parameters (in proper distance) are low-surface brightness objects. The proper distance selection method yields lower mass galaxies than the virial radius selection, as the larger virial radii of more massive galaxies in the field bring them closer in terms of $r/r_{\rm vir}$. Indeed, 6/9 absorbers yield larger impact parameters to their associated galaxies under the virial-radius selection method than those chosen by proper distance (by factors of $\gtrsim 2$ in some cases). This result accentuates the dilemma in attributing absorbers to individual nearby galaxies: selecting by virial radius may provide a physically motivated connection but may overlook important processes plausibly occurring, such as outflows or mass loss from dwarf galaxies.  On the other hand, galaxies increase in density with decreasing mass, and the mere proximity of a low-mass dwarf does not necessarily implicate it as the source of the gas, especially if both the absorber \emph{and} the dwarf galaxy may reside within the dark matter halo of a more massive galaxy.  In fact, for all six absorbers that have two different galaxies associated according to the two selection methods, \textit{the less massive galaxy falls within $1$ projected \rvir\ of the more massive galaxy}. 

Even given the ubiquity of dwarf galaxies in close proximity to QSO sightlines, the mass dependence of \cfour\ detections exhibited in Figure \ref{fig:masscol_rvir} calls into question the relative importance of the fainter dwarfs when proceeding to cross-correlate the CGM gas and galaxy properties.  For example, correlating the CGM absorption with the star formation activity of the massive galaxies selected by virial radius may mask the effects of the fainter star-forming galaxies within the same halo, such as if the massive galaxies' CGM are partly enriched by the dwarf galaxies residing therein.  In fact, it may be an important clue that the \cfour\ absorbers typically lie outside the putative virial radii of the nearest dwarf galaxy as the gas can more easily escape the dwarf's shallower potential well than that of the central galaxy, and the transfer of the metal-enriched gas outside the dwarf may even be aided by the presence of the counterpart \citep{Johnson:2015xy}.

Two important points arise from the result presented in Section \ref{sec:massciv}: (1) Isolated galaxies with $M_* < 9.5~\msuneq$ ($L\sim0.1 L*$) have a low covering fraction for $\ncfoureq > 10^{13.5} \cmt$ absorbers. (2) Dwarf galaxies, however, are frequently found at small impact parameters to sightlines/redshifts where \cfour\ is detected.  While this second point is itself not surprising, we now ask if our blind \cfour\ absorption has picked out special configurations of satellites and centrals such that the absorbers are tracing halos that are not only sufficiently massive but that also contain dwarf satellites at especially small impact parameters.  To address this question, we ran a simple Monte Carlo experiment to assess the typical distribution of impact parameters to faint dwarf galaxies given the presence of more massive galaxies within $\sim 1$ \rvir.  Within the SDSS footprint, we chose random coordinates to represent QSO sightlines and then a redshift at $z\leq0.015$ to localize simulated absorbers..  Then, we recorded the impact parameters and $r$-band magnitudes of all galaxies around the sightline within $|\delta z| < 0.002$.  After repeating this 50,000 times, we identified cases where sightlines passed within 180 kpc, 250 kpc, and 350 kpc of an $L > 0.1~L*$ galaxy, the approximate onset of \cfour\ absorption as a function of galaxy mass.  These impact parameters were chosen to represent the approximate virial radii of $L\sim0.1~L*$, $L\sim L*$, and $L>L*$.  Then, in each of these cases, the impact parameter to the nearest $L < 0.1~L*$ galaxy was tabulated.  

\begin{figure}[t!]
\centering
\includegraphics[width=.42\paperwidth]{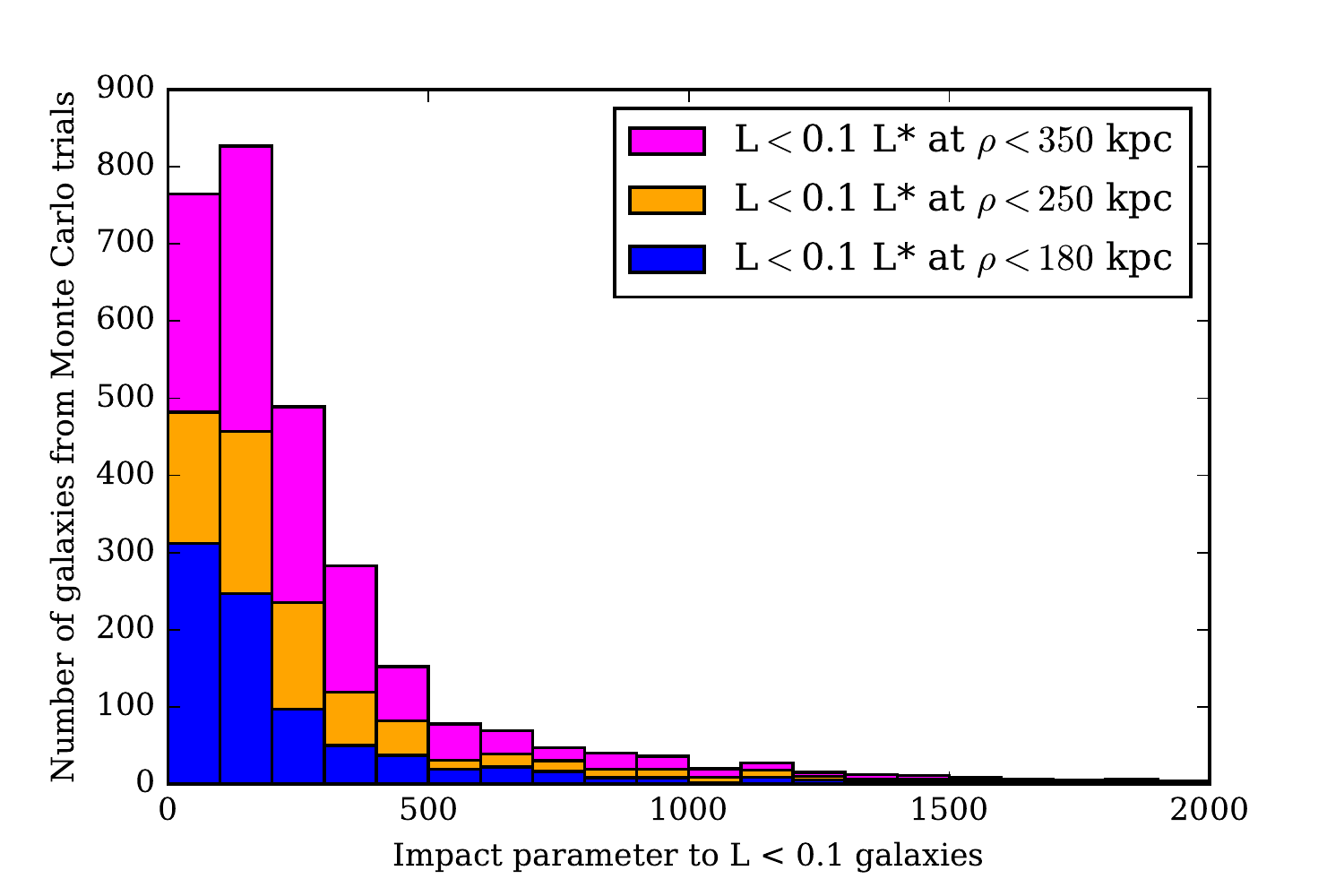}
\caption{Monte Carlo simulation results of the distribution of impact parameters of random `sightlines' and redshifts to faint dwarf galaxies $L < 0.1 L*$ when a more massive galaxy lies within three different impact parameters as indicated in the legend.  The three selections of impact parameters roughly correspond to the virial radii of $L\sim0.1L*$, $L\sim L*$, and $L>L*$ galaxies.  }
\label{fig:montecarlorhos}
\end{figure}

\begin{figure}[t!]
\centering
\includegraphics[width=.42\paperwidth]{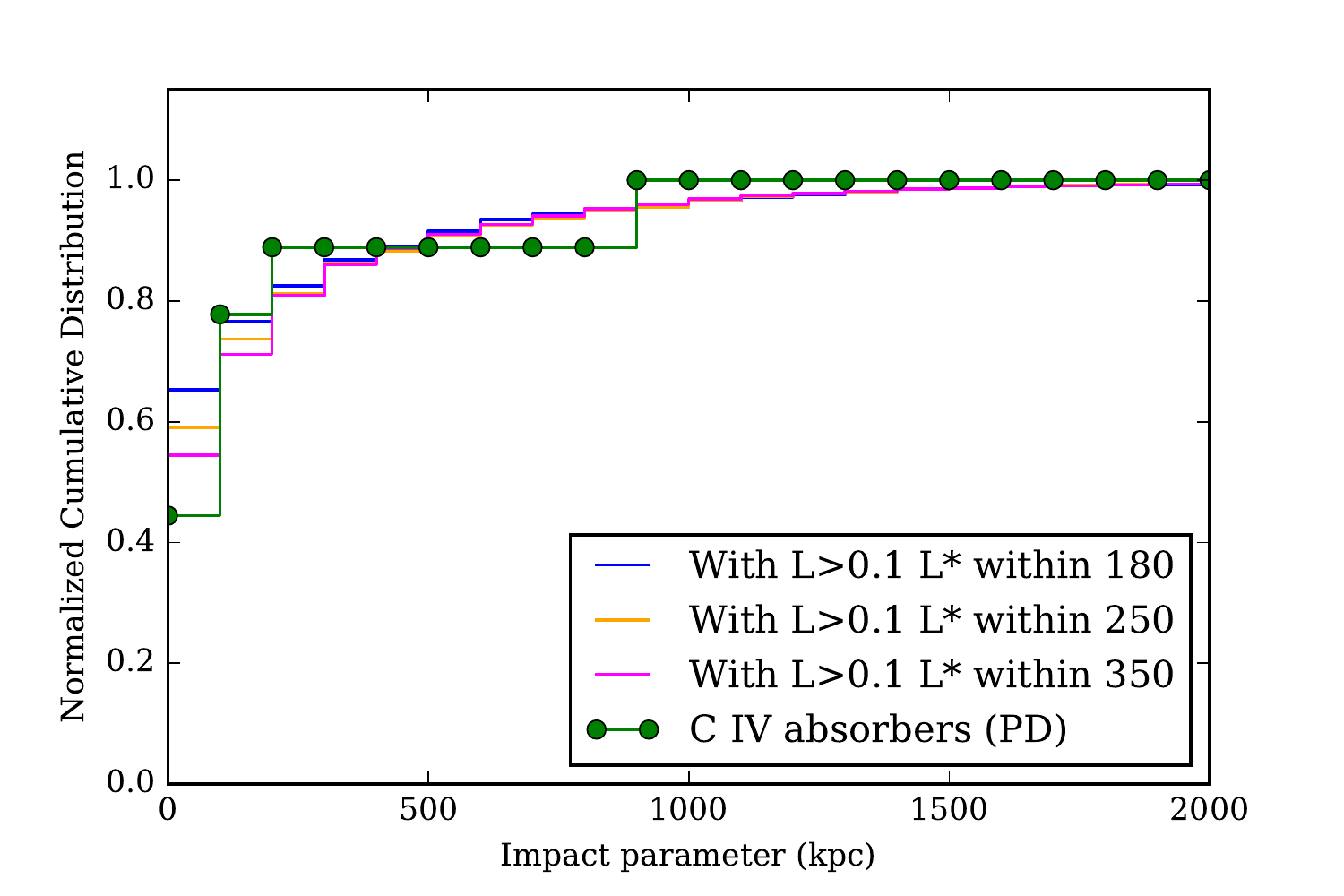}
\caption{The cumulative distribution of impact parameters to $L<0.1L*$ galaxies for the three scenarios presented in Figure \ref{fig:montecarlorhos} (in blue, orange, and magenta) alongside the observed impact parameter distribution from our proper distance-selected galaxy/absorber associations (green).  The Monte Carlo experiment suggests that configurations of galaxies and faint dwarf satellites associated with our blindly detected \cfour\ absorbers are similar to those that would be obtained given random sightlines and absorber redshifts. }
\label{fig:montecarlocum}
\end{figure}

Figure \ref{fig:montecarlorhos} shows the resulting impact parameter distributions; the color of each histogram signifies the impact parameter selection criterion for the more massive galaxy.  If our \cfour\ absorber-selected galaxy environments differed from those probed at random sightlines/redshifts, we might expect our distribution of proper-distance selected impact parameters to peak at smaller values than those from the Monte Carlo experiment. While some scenarios yield $L < 0.1~L*$ at much larger impact parameters than the $L > 0.1~L*$ galaxies, the less massive galaxies are overwhelmingly found nearer the sightline.  Figure \ref{fig:montecarlocum} shows the cumulative distribution of impact parameters to faint dwarfs $L<0.1~L*$ given each impact parameter selection for $L>0.1L*$ galaxies.  While our low-$z$ absorber sample is small, the Monte Carlo experiment reveals no strong evidence that the combination of faint dwarfs at small impact parameters with more massive galaxies at small $r/\rvireq$ are unique to the presence of \cfour\ absorbers.

To conclude this discussion, we emphasize the following: (1) Galaxy surveys around QSO sightlines that are complete down to even $L\sim 0.1~L*$ are in general missing fainter galaxies at small impact parameters $\rho \leq 100$ kpc. (2) However, these $L \lesssim 0.1~L*$ galaxies in isolation do not give rise to $\ncfoureq > 10^{13.5} \cmt$ absorbers.  The same is not true for \hone\ absorption, as seen in Figure \ref{fig:masscol_rvir}.  While the dwarf galaxies are believed to be effective at enriching the CGM/IGM on large scales \citep{Shen:2013lr}, the properties of \cfour\ absorption exhibited here may be more indicative of physical/dynamical conditions in moderately populated halos ($\mhaloeq \gtrsim 10^{11.5} \msuneq$).

\subsection{Comparing absorber statistics with previous surveys}
\label{sec:comparison}
As remarked in Section \ref{sec:c4colprofiles}, our \cfour\ column density profiles show some differences from those previously reported, such as the incidence of \cfour\ absorbers out to and beyond 1 \rvir.  Between the maximum impact parameter for a COS-Dwarfs \cfour\ detection ($\rho=0.66~r_{vir}$) and the maximum they probed ($\rho=1.33~r_{vir}$),  we report 2 of out of a possible 19 \cfour\ detections ($10^{+9}_{-5}$\%) compared to 0/11 in COS-Dwarfs ($0^{+8}_{-0}$\%), not a statistically significant difference.  However, our covering fraction is statistically nonzero with $>99.7\%$ confidence.  We also note the differences in galaxy masses between this work and COS-Dwarfs: our two detections in this impact parameter range occur at $0.73~r_{vir}$ from a $10^{10.4}~M_*$ galaxy,which exceeds the COS-Dwarfs target mass range, and at $1.27~r_{vir}$ from a $10^{9.9}~M_*$ galaxy, similar to the highest-mass COS-Dwarfs galaxies.  A similar result to that of COS-Dwarfs was reported by \citet{Liang:2014kx}, who detect no \cfour\ absorption at $\rho>0.7$ \rvir, but their galaxy sample has no imposed upper mass limit. Our detection fractions at $\rho<0.6$ \rvir\ and that of \citet{Liang:2014kx} are within the uncertainties of each other, but our detections at $\rho \geq 1$ \rvir\ are somewhat surprising given their lack of detections beyond $\rho>0.7$ \rvir; this could be a result of small-number statistics and differences in our halo radius/\rvir\ definitions. This rarer high-$\rho$/\rvir\ population of absorbers may offer key insights to outflow and stripping mechanisms that transport enriched gas out of galaxies and into the IGM.

Strong \cfour~absorption is patchy in the CGM, as shown above and also reported by \citet{Bordoloi:2014lr}, \citet{Borthakur:2013lr}, and others.  For star-forming galaxies alone, we only detect \cfour~within 150 kpc of 5/17 galaxies.  With a larger sample, COS-Dwarfs finds a \cfour~covering fraction that is  $\sim0.8$ in the inner 50 kpc of star-forming galaxies, but decreases to $<0.2$ in the outer 50 kpc of star-forming galaxies.  This characteristic of \cfour~absorption contrasts with the behavior seen in another strong-transition metal ion: \osix.  The COS-Halos team finds strong \osix~to be nearly ubiquitous in the CGM of their star-forming galaxy sample \citep{Tumlinson:2011kx}.  Their survey targeted QSO sightlines out to 150 kpc of $\sim L*$ galaxies with a redshift distribution chosen to provide coverage of the \osix~\lam\lam~1032,1028 doublet in the COS G130M grating.

We emphasize that \osix~and \cfour~are generally not simultaneously covered by our observations except for a narrow swath of redshift space, $0.11<z_{abs}<0.16$.  The lower-redshift end of \osix~coverage lies well beyond the cutoff for our low-z galaxy/absorber sample chosen for completeness to faint dwarf galaxies.  Therefore, we are prevented from directly comparing the \cfour~and \osix~characteristics within our galaxy/absorber pairs.  We must thus rely on \osix\ studies for this analysis. \citet{Bordoloi:2014lr} suggest that the differences between observations of \cfour~and \osix~may be attributable to the masses of the target galaxies (sub-$L*$ and $\sim L*$ for COS-Dwarfs and COS-Halos, respectively).  This conjecture appears to be supported by \citet{Prochaska:2011yq}, who find that the \osix~covering fraction is higher within the virialized halos of $L > 0.1~L*$ galaxies than for their dwarf galaxy counterparts ($L < 0.1~L*$).  This explanation is also consistent with results from blind \osix\  surveys \citep[e.g,][]{Tripp:2008lr}, which typically find much lower \osix\ column densities than those measured in the COS-Halos survey.  It is also possible that the COS-Halos sample probes more rare luminous galaxies that have unusually high quantities of \osix\  in their halos.   This as-yet unexplored \cfour-\osix\ overlap region at $z>0.11$ (at least in terms of HST/COS capability) holds great promise for investigating the apparent discrepancy between the profiles of these two ions around galaxies.

\subsection{Group/cluster environments and the CGM}
\label{sec:groupclusterdiscuss}
A novel facet of this work is the investigation of environmental effects on the metal-enriched gas traced by \cfour.  We have found an apparent suppression of circumgalactic \cfour\ absorbers within halos of log $\mhaloeq = 12\sim13~\msuneq$, or approximately the transition region where halos that typically host isolated galaxies transition to hosting groups \citep{Yang:2003yq}.  \citet{Oppenheimer:2016lr} contend that the prevalence of \osix\ and the correlation between sSFR and N(\osix) found by COS-Halos can be attributed to the similarity between the virial temperature corresponding to dark matter halos hosting $L*$ galaxies and the peak temperature for \osix\ abundances under collisional ionization. 

The peak temperature for \cfour\ due to collisional ionization equilibrium occurs at $\sim 10^{5.1}$ K \citep{Gnat:2007fk,Verner:1994fk}.  Using the expression given by \citet{Oppenheimer:2016lr} for the virial temperature of a dark matter halo, which they find follows the typical gas temperature for halos in their simulations, the peak temperature for \cfour\ corresponds to log $\mhaloeq = 11.1 \msuneq$.  We find that the CGM \cfour\ detection rate increases sharply at log $\mstareq = 9.0-9.5~\msuneq$ (cf., Figure \ref{fig:masscol_rvir}), which corresponds to halo masses of log $\mhaloeq = 11.0-11.2~\msuneq$ from halo abundance matching.  Thus, the gas we observe via \cfour\ may in part be collisionally ionized. The simulations of \citet{Cen:2011zl} indicate that \cfour\ absorbers are dominated by photoionization after $z=3$, but some contribution of absorbers with \cfourcol = $10^{12-14}~ \cmt$ does arise from gas at temperatures near $\sim 10^{5.1}$ K.

However, it is also clear from Figure \ref{fig:masscol_rvir} that \cfour\ absorption persists when $\mstareq > 10.0~\msuneq$, which corresponds to $\mhaloeq > 11.5~\msuneq$ from halo abundance matching.  The virial temperatures of these halos well exceed the peak of the triply ionized state of carbon, and this simple picture does not quite reconcile the results of Sections \ref{sec:massciv} and \ref{sec:envirociv}.  Section \ref{sec:envirociv} shows that the \cfour\ detection rate indeed declines just beyond this halo mass, and \hone\ persists beyond the halo mass range where \cfour\ is deficient.  The overdense region may simply contain more gas, compensating for the lowered abundance of the \cfour\ ionization state when the dominant state has transitioned to C~\textsc{v}.  \citet{Schaye:2003rr} found that the observed \cfour\ at $1.8 < z < 4.1$ must arise from photoionization based on C~\textsc{iii}/\cfour\ ratios.  Unfortunately, we do not cover C~\textsc{iii} ($\lambda_r = 977$\AA) for any of the galaxy/group-absorber pairs in this study to use this diagnostic.

\begin{figure}[t!]
\centering
\includegraphics[width=.42\paperwidth]{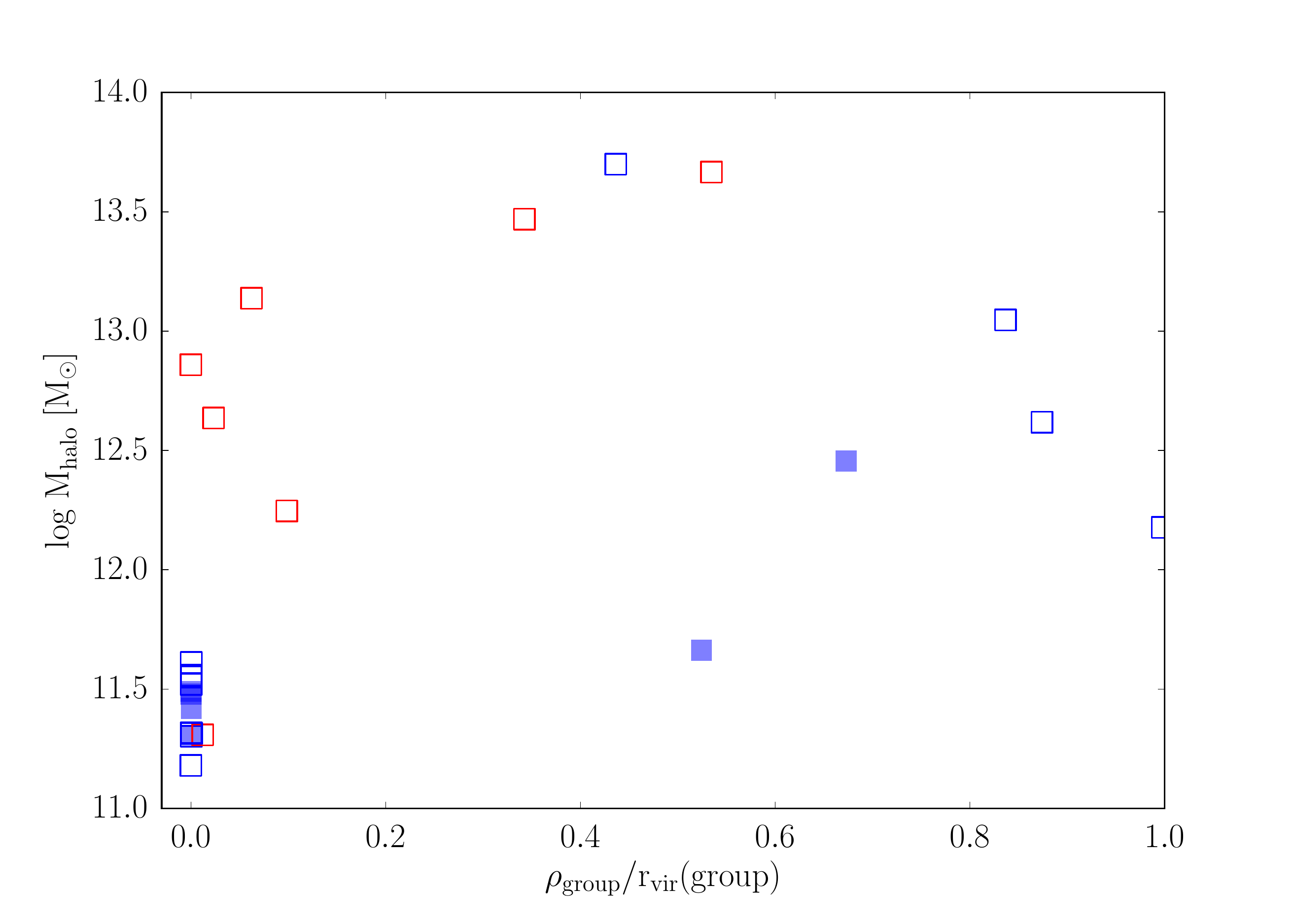}
\caption{The halo masses of our \cfour\ galaxy/absorber environment sample versus impact parameter relative to the group virial radius.  Open squares represent \cfour\ nondetections, and filled squares denote detections.  We note that the lone detection of \cfour\ in  an $\mhaloeq > 10^{12}~\msuneq$ occurs beyond 0.5 \rvir in projection.  Filling in this parameter space will provide insight to the effects of environment on the CGM (and presumably the host galaxies themselves).}
\label{fig:halomassrvir}
\end{figure}

Lastly, we suggest that a galaxy's CGM may well be affected by its relationship to the encompassing group.  The galaxy associated with the one detection of \cfour\ shown in Figure \ref{fig:c4halomass} where $\mhaloeq>12~\msuneq$ lies at $r > 0.5~\rvireq$ relative to the \emph{group} virial radius.   No detection occurs for galaxies within $r = 0.5~\rvireq$(group) of halos with $\mhaloeq>12~\msuneq$. In Figure \ref{fig:halomassrvir}, we plot the halo mass of each galaxy/absorber from Figure \ref{fig:c4halomass} against the projected distance of the galaxy relative to the associated group virial radius; the solid squares correspond to \cfour\ detections and the open squares nondetections. This representation of our data underscores that large regions of parameter space pertaining to the CGM/environment connection remain unexplored.  Many points are clustered in the lower-left of this plot, and these largely correspond to isolated galaxies.  If larger samples filling in the large $\rho_{\rm group}/{\rm r}_{\rm vir}{\rm (group)}$ but high-\mhalo\ regime reveal little \cfour\ absorption, these galaxies' CGM may indeed be undergoing ram-pressure or tidal stripping.   

\section{Summary and Conclusions}
\label{sec:summary}
This work, Paper III of an ongoing program, presents initial survey results employing public archival galaxy survey data (SDSS, RC3, etc.), including spectroscopy and imaging, covering fields around the QSO sightlines employed in our blind \cfour\ survey presented in Paper II.  To achieve high completeness to faint dwarf galaxies ($L \sim 0.01 L*$), we limit our analyses of individual galaxy-absorber associations (Section \ref{sec:indivgals})  to $z\leq0.015$ based on the magnitude limit of the SDSS spectroscopic data.   Our \cfour-environment analysis (Section \ref{sec:envirociv}) includes galaxies and absorbers to $z<0.055$ building on the increased detection rate of \cfour\ in the CGM of $M_* \gtrsim 10^{9.5} \msuneq$ host galaxies.

We summarize our main results below:

\begin{enumerate}

\item Selecting individual galaxies that are associated with \cfour\ absorbers is generally ambiguous.  We have selected associated galaxies based both on their proper distance separations and in terms of their virial radii.  For 6/9 of our blindly detected \cfour\ absorbers at $z\leq0.015$, these differing criteria yield associations with different galaxies: the galaxies with the smallest proper distance impact parameters are dwarf galaxies, most with low surface brightnesses.

\item When selecting galaxy/absorber associations by proper distance, we find galaxies at impact parameters $\rho < 200$ kpc with the exception of one galaxy/absorber pair with $\rho >350$ kpc. This large-$\rho$ absorber also has a velocity separation $\delta v >700$ km/s. In addition, we observe a prevalence of \hone\ out to 350 kpc; however, at least half of the $\rho>300$~kpc \lya\ absorbers have low column densities that are similar to those statistically unassociated with galaxies.  When selecting by virial radius, 3/9 \cfour\ absorbers fall $\gtrsim 1~\rvireq$ of any galaxy detected.

\item We find that galaxies with $M_* > 10^{9.5} M_{\odot}$ show a significantly greater \cfour\ detection rate within 1~\rvir\ than galaxies of lower mass.  This mass dependence does not extend to \hone, as high detection rates occur for \hone\ within 1 \rvir\ of galaxies with masses $M_* < 10^{8} M_{\odot}$. 

\item At $z<0.055$, we do not detect \cfour\ within 160 kpc of any galaxy residing in an environment containing more than seven $\mathcal{M}_r \leq -19$ galaxies within 1.5 Mpc but find a 57\% detection rate within 160 kpc of galaxies when six or fewer galaxies of this luminosity reside within 1.5 Mpc.  It is unclear what mechanisms lead to the dearth of \cfour\ absorbers in the inner CGM in dense environments, but these regions are not devoid of gas, as we detect \hone\ independent of environment.  When using group halo masses rather than galaxy counts to quantify the environments of these same galaxies, the \cfour\ detection rate falls to 0\% at $M_{h}>10^{12.7} M_{\odot}$.

\end{enumerate}

The data presented here highlight the close association between galaxies and metal-enriched gas using a novel combination of sensitivity to low-luminosity galaxies and blindly discovered \cfour\ absorbers.  These findings emphasize that the CGM is a key constituent in galaxy ecosystems, reflecting the buildup of enriched material through mass assembly while signifying the dependence of physical conditions on large-scale environment.

\section*{Acknowledgements}
We thank John Stocke, Hsiao-Wen Chen, Houjun Mo, Kate Rubin, Romeel Dav\'e, Martin Weinberg, Andrew Battisti, Brice M\'enard, and Sandy Faber for helpful discussions at various stages of survey design and analysis.   Support for this research was provided by NASA through grants HST-GO-11741, HST-GO-11598, HST-GO-12248, and HST-AR-13894 from the Space Telescope Science Institute, which is operated by the Association of Universities for Research in Astronomy, Incorporated, under NASA contract NAS5-26555.  Funding for SDSS-III has been provided by the Alfred P. Sloan Foundation, the Participating Institutions, the National Science Foundation, and the U.S. Department of Energy Office of Science. The SDSS-III web site is http://www.sdss3.org/.

SDSS-III is managed by the Astrophysical Research Consortium for the Participating Institutions of the SDSS-III Collaboration including the University of Arizona, the Brazilian Participation Group, Brookhaven National Laboratory, Carnegie Mellon University, University of Florida, the French Participation Group, the German Participation Group, Harvard University, the Instituto de Astrofisica de Canarias, the Michigan State/Notre Dame/JINA Participation Group, Johns Hopkins University, Lawrence Berkeley National Laboratory, Max Planck Institute for Astrophysics, Max Planck Institute for Extraterrestrial Physics, New Mexico State University, New York University, Ohio State University, Pennsylvania State University, University of Portsmouth, Princeton University, the Spanish Participation Group, University of Tokyo, University of Utah, Vanderbilt University, University of Virginia, University of Washington, and Yale University.

\section{Appendix}
\begin{figure*}[h!]
\centering
\includegraphics[width=0.85\paperwidth]{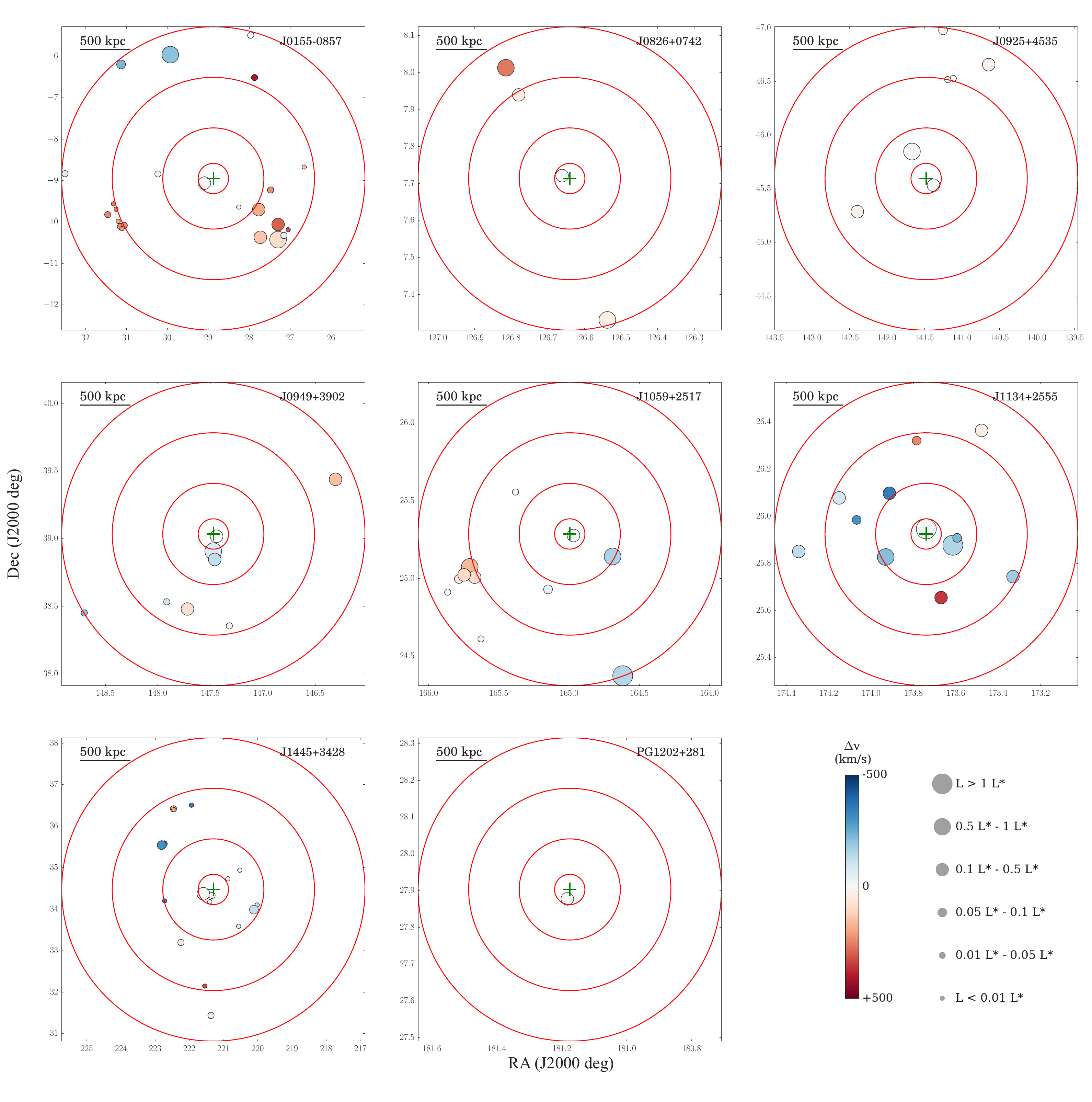}
\caption{The galaxy environments of the \cfour~detections plotted in Fig. \ref{fig:gals1000kpc}.  The crosses in the center of each map mark the QSO sightline, and the concentric circles indicate impact parameters of 150, 500, 1000, and 1500 kpc. The sizes of the symbols depict the galaxy luminosities and the color indicates the velocity offset from the detected absorber as indicated by the colorbar.}
\label{fig:densdet}
\end{figure*}

Here, we present maps of the galaxy populations around the QSO sightlines employed in our galaxy environment/absorber analysis.  Figures \ref{fig:densdet}-\ref{fig:densnondet} show the galaxies composing the environments that are plotted as individual points in Figure \ref{fig:gals1000kpc}; these maps reveal the contrast in galaxy density that result in the differing detection rates of \cfour\ absorption.  Figures \ref{fig:densdet}-\ref{fig:densnondet} include only $\mathcal{M}_r \leq -19$ galaxies and therefore are not directly comparable to Figure \ref{fig:absmaps}, which uses no minimum luminosity.  For a different perspective, Figure \ref{fig:wedgie} shows the galaxy impact parameters within 1.5 Mpc of the QSO sightlines depicted in Figure \ref{fig:gals1000kpc}.  Here, the galaxy overdensities in large-scale structures probed by the QSO sightlines are conspicuous at various redshifts.  The redshifts of the $\rho \leq 160$ kpc galaxies are marked with vertical solid and dashed lines for \cfour\ detections and non-detections, respectively.

\begin{figure*}[t!]
\centering
\includegraphics[width=0.85\paperwidth]{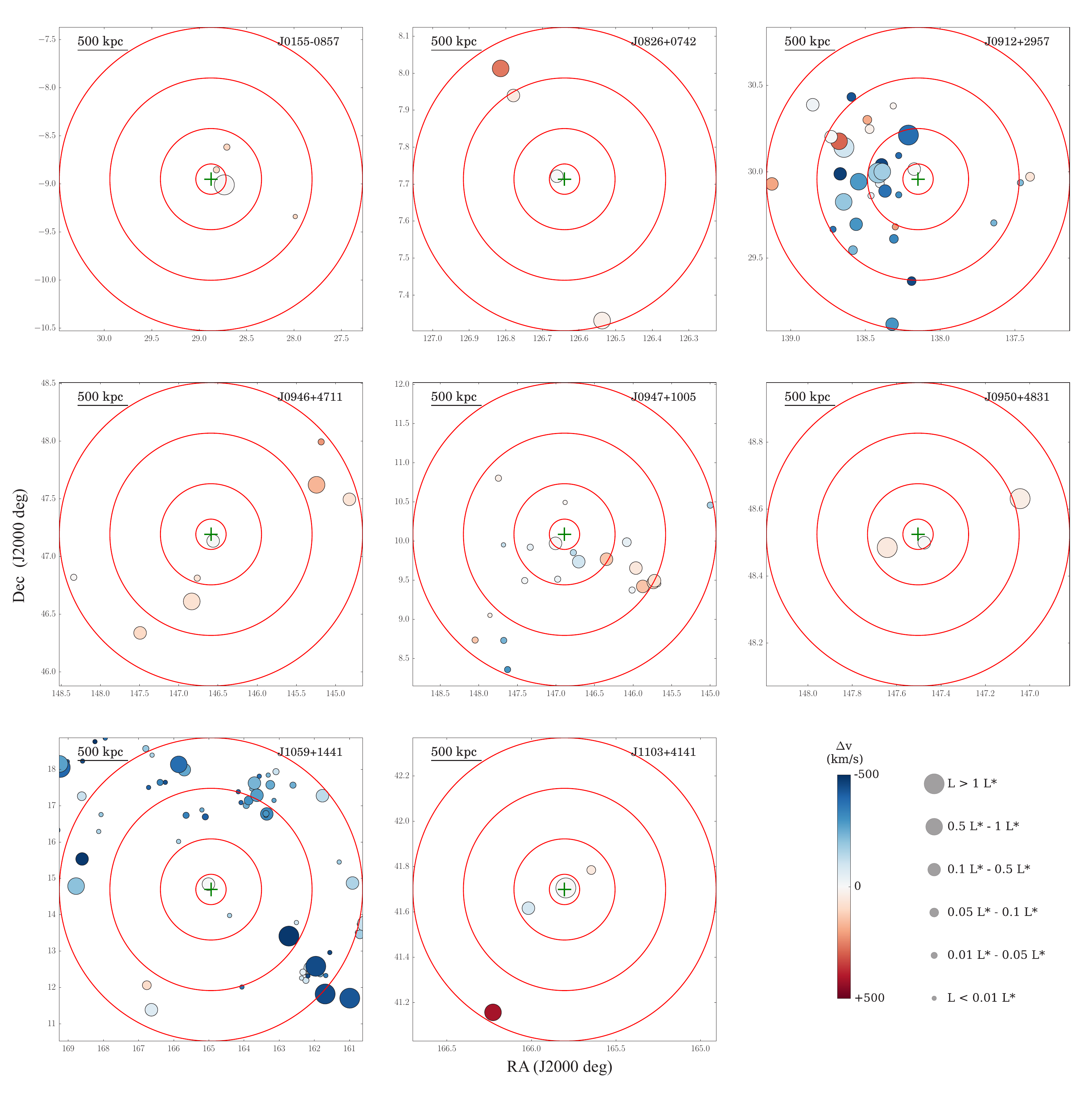}
\caption{Same as Fig. \ref{fig:densdet} but for \cfour~\textit{nondetections}.  The velocity offsets are calculated from the innermost galaxy.}
\label{fig:densnondet}
\end{figure*}

\setcounter{figure}{\value{figure}-1}
\begin{figure*}[t!]
\centering
\includegraphics[width=0.85\paperwidth]{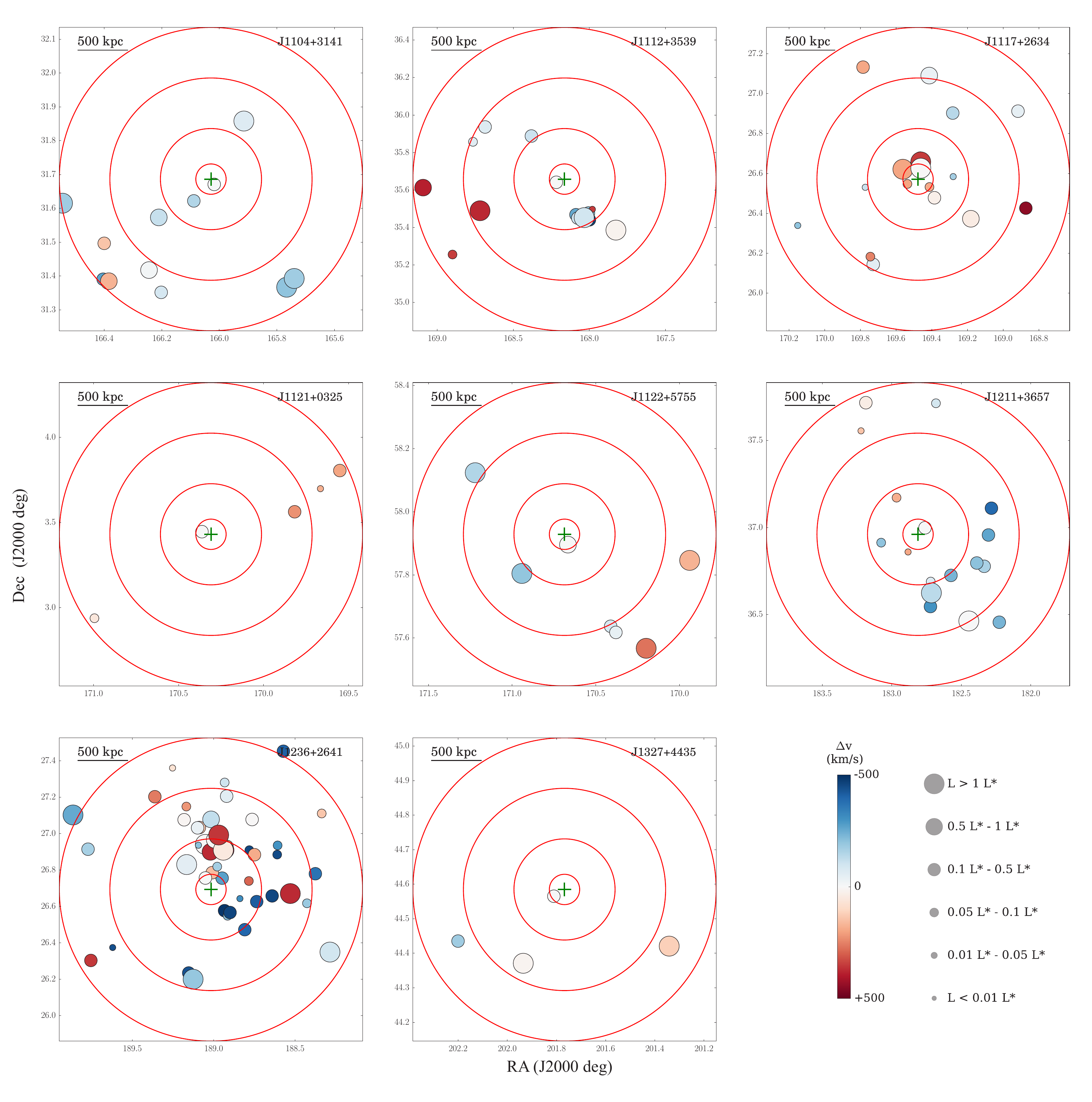}
\caption{Same as Fig. \ref{fig:densdet} but for \cfour~\textit{nondetections}.  The velocity offsets are calculated from the innermost galaxy.}
\label{fig:densnondet}
\end{figure*}

\setcounter{figure}{\value{figure}-1}
\begin{figure*}[t!]
\centering
\includegraphics[width=0.85\paperwidth]{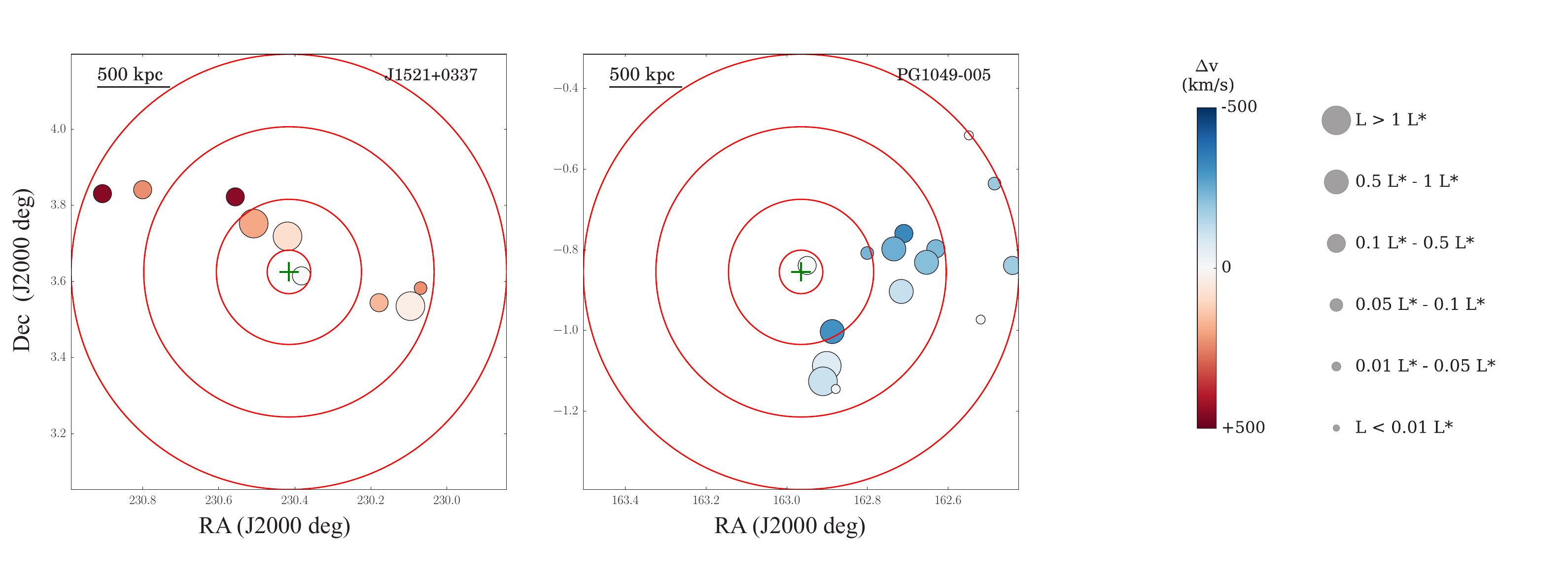}
\caption{continued.}
\end{figure*}

\begin{figure*}[t!]
\centering
\includegraphics[width=0.82\paperwidth]{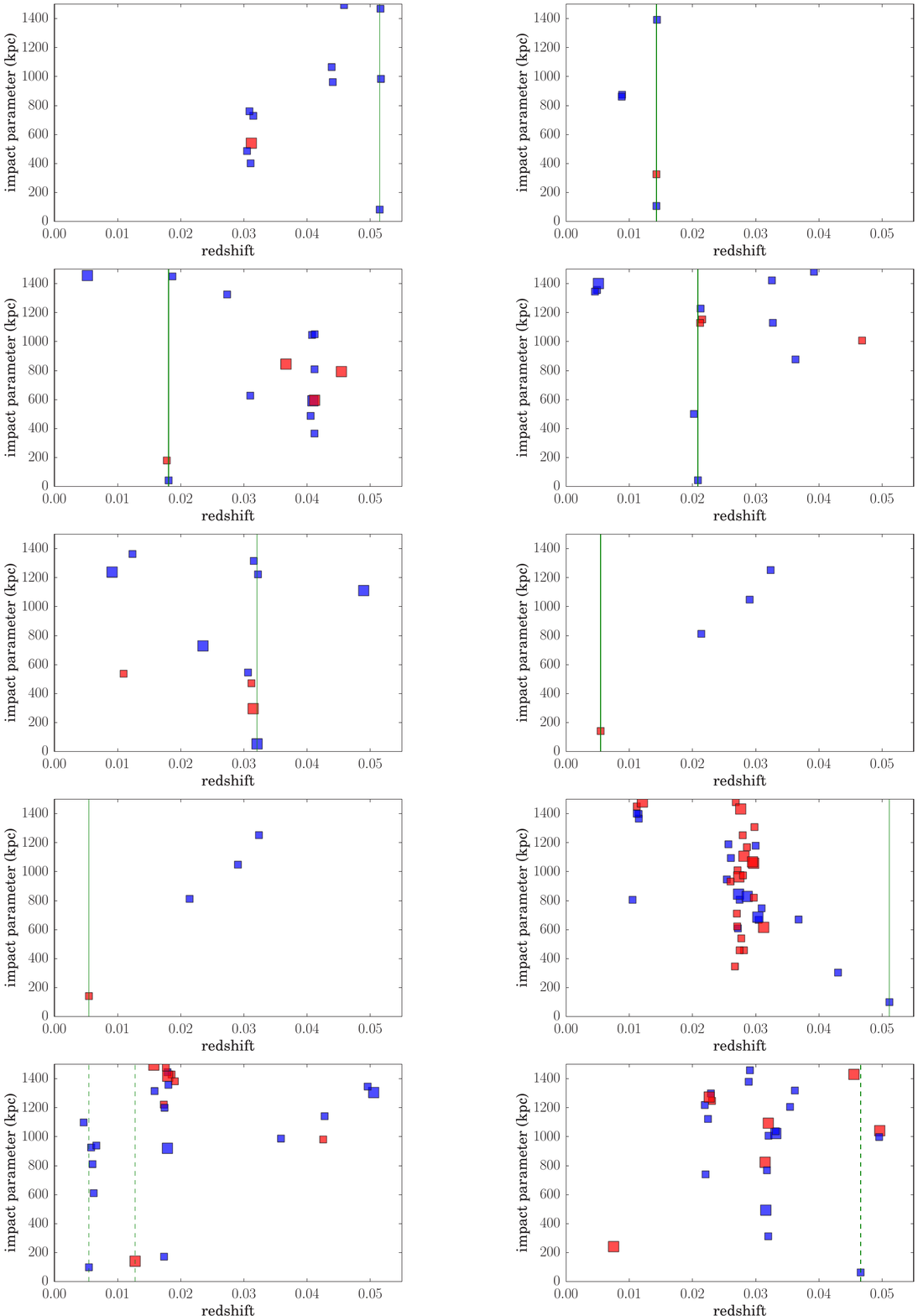}
\caption{Impact parameter vs. redshift for galaxies within 1500 kpc of QSO sightlines represented in Fig. \ref{fig:gals1000kpc} and with maps in Figs. \ref{fig:densdet} and \ref{fig:densnondet}.  Vertical lines appear at redshifts where an $\mathcal{M}_r \leq -19$ galaxy falls within 160 kpc of the sightline; solid lines mark \cfour~detections, and dashed mark \cfour~nondetections.  }
\label{fig:wedgie}
\end{figure*}

\setcounter{figure}{\value{figure}-1}
\begin{figure*}[htb!]
\centering
\includegraphics[width=0.82\paperwidth]{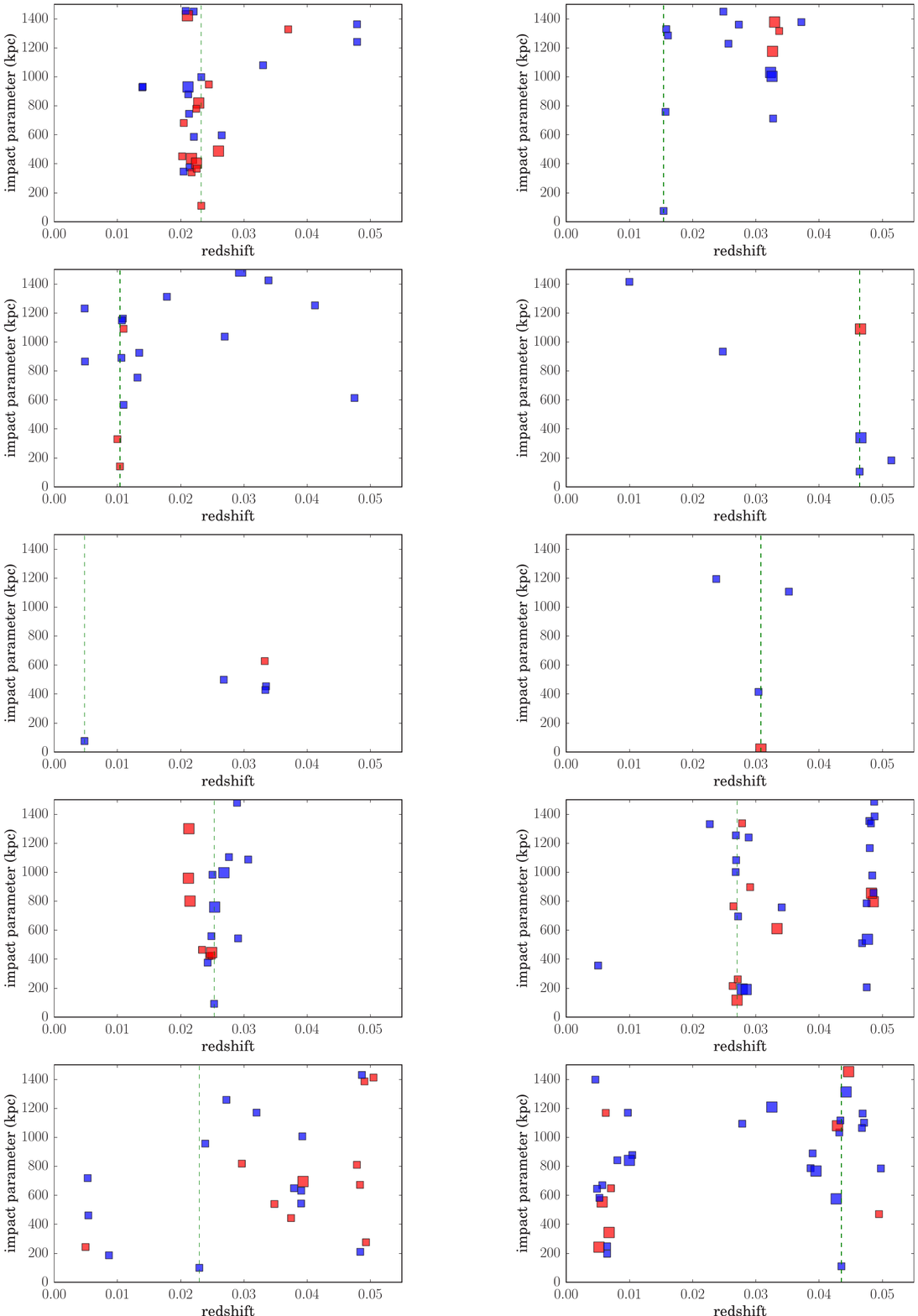}
\caption{continued. }
\end{figure*}

\setcounter{figure}{\value{figure}-1}
\begin{figure*}[htb!]
\centering
\includegraphics[width=0.82\paperwidth]{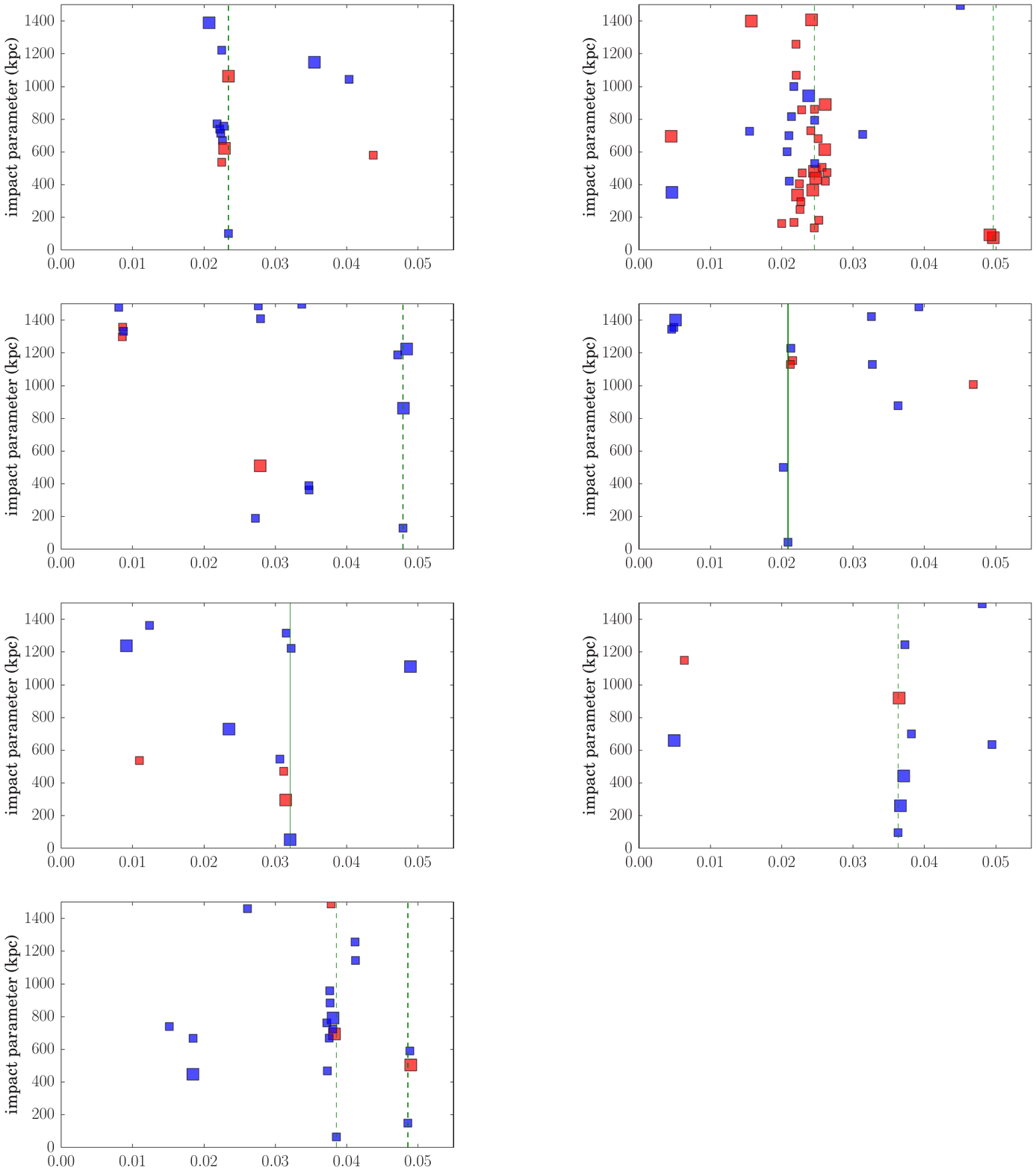}
\caption{continued. }
\end{figure*}


\begin{thebibliography}{90}
\expandafter\ifx\csname natexlab\endcsname\relax\def\natexlab#1{#1}\fi

\bibitem[{Abell(1965)}]{Abell:1965fk}
Abell, G.~O. 1965, ARAA, 3, 1

\bibitem[{Alam {et~al.}(2015)Alam, Albareti, Prieto, Anders, Anderson, Andrews,
  Armengaud, Aubourg, Bailey, Bautista, Beaton, Beers, Bender, Berlind,
  Beutler, Bhardwaj, Bird, Bizyaev, Blake, Blanton, Blomqvist, Bochanski,
  Bolton, Bovy, Bradley, Brandt, Brauer, Brinkmann, Brown, Brownstein, Burden,
  Burtin, Busca, Cai, Capozzi, Rosell, Carrera, Chen, Chiappini, Chojnowski,
  Chuang, Clerc, Comparat, Covey, Croft, Cuesta, Cunha, da~Costa, Da~Rio,
  Davenport, Dawson, De~Lee, Delubac, Deshpande, Dutra-Ferreira, Dwelly, Ealet,
  Ebelke, Edmondson, Eisenstein, Escoffier, Esposito, Fan, Fern{\'a}ndez-Alvar,
  Feuillet, Ak, Finley, Finoguenov, Flaherty, Fleming, Font-Ribera, Foster,
  Frinchaboy, Galbraith-Frew, Garc{\'\i}a-Hern{\'a}ndez, P{\'e}rez, Gaulme, Ge,
  G{\'e}nova-Santos, Ghezzi, Gillespie, Girardi, Goddard, Gontcho,
  Hern{\'a}ndez, Grebel, Grieb, Grieves, Gunn, Guo, Harding, Hasselquist,
  Hawley, Hayden, Hearty, Ho, Hogg, Holley-Bockelmann, Holtzman, Honscheid,
  Huehnerhoff, Jiang, Johnson, Kinemuchi, Kirkby, Kitaura, Klaene, Kneib,
  Koenig, Lam, Lan, Lang, Laurent, Goff, Leauthaud, Lee, Lee, Licquia, Liu,
  Long, L{\'o}pez-Corredoira, Lorenzo-Oliveira, Lucatello, Lundgren, Lupton,
  Mack~III, Mahadevan, Maia, Majewski, Malanushenko, Malanushenko, Manchado,
  Manera, Mao, Maraston, Marchwinski, Margala, Martell, Martig, Masters,
  McBride, McGehee, McGreer, McMahon, M{\'e}nard, Menzel, Merloni,
  M{\'e}sz{\'a}ros, Miller, Miralda-Escud{\'e}, Miyatake, Montero-Dorta, More,
  Morice-Atkinson, Morrison, Muna, Myers, Newman, Neyrinck, Nguyen, Nichol,
  Nidever, Noterdaeme, Nuza, O'Connell, O'Connell, O'Connell, Ogando, Olmstead,
  Oravetz, Oravetz, Osumi, Owen, Padgett, Padmanabhan, Paegert,
  Palanque-Delabrouille, Pan, Parejko, Park, P{\^a}ris, Pattarakijwanich,
  Pellejero-Ibanez, Pepper, Percival, P{\'e}rez-Fournon, P{\'e}rez-R{\`a}fols,
  Petitjean, Pieri, Pinsonneault, de~Mello, Prada, Prakash, Price-Whelan,
  Raddick, Rahman, Reid, Rich, Rix, Robin, Rockosi, Rodrigues,
  Rodr{\'\i}guez-Rottes, Roe, Ross, Ross, Rossi, Ruan,
  Rubi{\~n}o-Mart{\textbackslash}'{\textbackslash}in, Rykoff, Salazar-Albornoz,
  Salvato, Samushia, S{\'a}nchez, Santiago, Sayres, Schiavon, Schlegel,
  Schmidt, Schneider, Schultheis, Schwope, Sc{\'o}ccola, Sellgren, Seo, Shane,
  Shen, Shetrone, Shu, Sivarani, Skrutskie, Slosar, Smith, Sobreira, Stassun,
  Steinmetz, Strauss, Streblyanska, Swanson, Tan, Tayar, Terrien, Thakar,
  Thomas, Thompson, Tinker, Tojeiro, Troup, Vargas-Maga{\~n}a, Vazquez, Verde,
  Viel, Vogt, Wake, Wang, Weaver, Weinberg, Weiner, White, Wilson, Wisniewski,
  Wood-Vasey, Y{\`e}che, York, Zakamska, Zamora, Zasowski, Zehavi, Zhao, Zheng,
  Zhou, Zhou, Zhu, \& Zou}]{Alam:2015kq}
Alam, S., Albareti, F.~D., Prieto, C.~A., {et~al.} 2015, The Astrophysical
  Journal Supplement Series, 219, 12, arXiv: 1501.00963

\bibitem[{Bahcall {et~al.}(1993)Bahcall, Bergeron, Boksenberg, Hartig, Jannuzi,
  Kirhakos, Sargent, Savage, Schneider, Turnshek, Weymann, \&
  Wolfe}]{Bahcall:1993zr}
Bahcall, J.~N., Bergeron, J., Boksenberg, A., {et~al.} 1993, ApJS, 87, 1

\bibitem[{Blanton \& Roweis(2007)}]{Blanton:2007ys}
Blanton, M.~R., \& Roweis, S. 2007, AJ, 133, 734

\bibitem[{Blanton {et~al.}(2003)Blanton, Hogg, Bahcall, Brinkmann, Britton,
  Connolly, Csabai, Fukugita, Loveday, Meiksin, Munn, Nichol, Okamura, Quinn,
  Schneider, Shimasaku, Strauss, Tegmark, Vogeley, \&
  Weinberg}]{Blanton:2003kx}
Blanton, M.~R., Hogg, D.~W., Bahcall, N.~A., {et~al.} 2003, ApJ, 592, 819

\bibitem[{Bordoloi {et~al.}(2011)Bordoloi, Lilly, Knobel, Bolzonella, Kampczyk,
  Carollo, Iovino, Zucca, Contini, Kneib, Le~Fevre, Mainieri, Renzini,
  Scodeggio, Zamorani, Balestra, Bardelli, Bongiorno, Caputi, Cucciati, de~la
  Torre, de~Ravel, Garilli, Kova{\v c}, Lamareille, Le~Borgne, Le~Brun, Maier,
  Mignoli, Pello, Peng, Perez~Montero, Presotto, Scarlata, Silverman, Tanaka,
  Tasca, Tresse, Vergani, Barnes, Cappi, Cimatti, Coppa, Diener, Franzetti,
  Koekemoer, L{\'o}pez-Sanjuan, {McCracken}, Moresco, Nair, Oesch, Pozzetti, \&
  Welikala}]{Bordoloi:2011uq}
Bordoloi, R., Lilly, S.~J., Knobel, C., {et~al.} 2011, ApJ, 743, 10

\bibitem[{Bordoloi {et~al.}(2014)Bordoloi, Tumlinson, Werk, Oppenheimer,
  Peeples, Prochaska, Tripp, Katz, Dav{\'e}, Fox, Thom, Ford, Weinberg,
  Burchett, \& Kollmeier}]{Bordoloi:2014lr}
Bordoloi, R., Tumlinson, J., Werk, J.~K., {et~al.} 2014, ApJ, 796, 136

\bibitem[{Borthakur {et~al.}(2013)Borthakur, Heckman, Strickland, Wild, \&
  Schiminovich}]{Borthakur:2013lr}
Borthakur, S., Heckman, T., Strickland, D., Wild, V., \& Schiminovich, D. 2013,
  ApJ, 768, 18

\bibitem[{Brinchmann {et~al.}(2004)Brinchmann, Charlot, White, Tremonti,
  Kauffmann, Heckman, \& Brinkmann}]{Brinchmann:2004lr}
Brinchmann, J., Charlot, S., White, S. D.~M., {et~al.} 2004, MNRAS, 351, 1151

\bibitem[{Burchett {et~al.}(2013)Burchett, Tripp, Werk, Howk, Prochaska, Ford,
  \& Dav{\'e}}]{Burchett:2013qy}
Burchett, J.~N., Tripp, T.~M., Werk, J.~K., {et~al.} 2013, ApJL, 779, L17

\bibitem[{Burchett {et~al.}(2015)Burchett, Tripp, Prochaska, Werk, Tumlinson,
  O'Meara, Bordoloi, Katz, \& Willmer}]{Burchett:2015rf}
Burchett, J.~N., Tripp, T.~M., Prochaska, J.~X., {et~al.} 2015, ApJ, 815, 91

\bibitem[{Butcher \& Oemler(1984)}]{Butcher:1984lr}
Butcher, H., \& Oemler, Jr., A. 1984, ApJ, 285, 426

\bibitem[{Cardelli {et~al.}(1989)Cardelli, Clayton, \&
  Mathis}]{Cardelli:1989lr}
Cardelli, J.~A., Clayton, G.~C., \& Mathis, J.~S. 1989, The Astrophysical
  Journal, 345, 245

\bibitem[{Cen \& Chisari(2011)}]{Cen:2011zl}
Cen, R., \& Chisari, N.~E. 2011, ApJ, 731, 11

\bibitem[{Chabrier(2003)}]{Chabrier:2003pd}
Chabrier, G. 2003, Publications of the Astronomical Society of the Pacific,
  115, 763

\bibitem[{Chen {et~al.}(2001)Chen, Lanzetta, \& Webb}]{Chen:2001lr}
Chen, H.-W., Lanzetta, K.~M., \& Webb, J.~K. 2001, ApJ, 556, 158

\bibitem[{Chen {et~al.}(2005)Chen, Prochaska, Weiner, Mulchaey, \&
  Williger}]{Chen:2005kx}
Chen, H.-W., Prochaska, J.~X., Weiner, B.~J., Mulchaey, J.~S., \& Williger,
  G.~M. 2005, ApJL, 629, L25

\bibitem[{Colless {et~al.}(2001)Colless, Dalton, Maddox, Sutherland, Norberg,
  Cole, Bland-Hawthorn, Bridges, Cannon, Collins, Couch, Cross, Deeley,
  De~Propris, Driver, Efstathiou, Ellis, Frenk, Glazebrook, Jackson, Lahav,
  Lewis, Lumsden, Madgwick, Peacock, Peterson, Price, Seaborne, \&
  Taylor}]{Colless:2001ys}
Colless, M., Dalton, G., Maddox, S., {et~al.} 2001, Monthly Notices of the
  Royal Astronomical Society, 328, 1039

\bibitem[{Corwin {et~al.}(1994)Corwin, Buta, \& de~Vaucouleurs}]{Corwin:1994vn}
Corwin, Jr., H.~G., Buta, R.~J., \& de~Vaucouleurs, G. 1994, The Astronomical
  Journal, 108, 2128

\bibitem[{Dav{\'e} {et~al.}(2011)Dav{\'e}, Oppenheimer, \&
  Finlator}]{Dave:2011fk}
Dav{\'e}, R., Oppenheimer, B.~D., \& Finlator, K. 2011, MNRAS, 415, 11

\bibitem[{Davies \& Lewis(1973)}]{Davies:1973lr}
Davies, R.~D., \& Lewis, B.~M. 1973, MNRAS, 165, 231

\bibitem[{de~Vaucouleurs {et~al.}(1991)de~Vaucouleurs, de~Vaucouleurs, Corwin,
  Buta, Paturel, \& Fouqu{\'e}}]{Vaucouleurs:1991rt}
de~Vaucouleurs, G., de~Vaucouleurs, A., Corwin, Jr., H.~G., {et~al.} 1991,
  Third {Reference} {Catalogue} of {Bright} {Galaxies}. {Volume} {I}:
  {Explanations} and references. {Volume} {II}: {Data} for galaxies between 0h
  and 12h. {Volume} {III}: {Data} for galaxies between 12h and 24h.

\bibitem[{Dekel \& Birnboim(2006)}]{Dekel:2006lr}
Dekel, A., \& Birnboim, Y. 2006, MNRAS, 368, 2

\bibitem[{Dressler(1980)}]{Dressler:1980qy}
Dressler, A. 1980, ApJ, 236, 351

\bibitem[{Finn {et~al.}(2016)Finn, Morris, Tejos, Crighton, Perry, Fumagalli,
  Bielby, Theuns, Schaye, Shanks, Liske, Gunawardhana, \& Bartle}]{Finn:2016fk}
Finn, C.~W., Morris, S.~L., Tejos, N., {et~al.} 2016, Monthly Notices of the
  Royal Astronomical Society, 460, 590

\bibitem[{Ford {et~al.}(2013)Ford, Oppenheimer, Dav{\'e}, Katz, Kollmeier, \&
  Weinberg}]{Ford:2013lr}
Ford, A.~B., Oppenheimer, B.~D., Dav{\'e}, R., {et~al.} 2013, MNRAS, 432, 89

\bibitem[{Fumagalli {et~al.}(2011)Fumagalli, Prochaska, Kasen, Dekel, Ceverino,
  \& Primack}]{Fumagalli:2011qy}
Fumagalli, M., Prochaska, J.~X., Kasen, D., {et~al.} 2011, MNRAS, 418, 1796

\bibitem[{Gnat \& Sternberg(2007)}]{Gnat:2007fk}
Gnat, O., \& Sternberg, A. 2007, The Astrophysical Journal Supplement Series,
  168, 213

\bibitem[{Gunn \& Gott(1972)}]{Gunn:1972qy}
Gunn, J.~E., \& Gott, III, J.~R. 1972, ApJ, 176, 1

\bibitem[{Haas {et~al.}(2012)Haas, Schaye, \& Jeeson-Daniel}]{Haas:2012kq}
Haas, M.~R., Schaye, J., \& Jeeson-Daniel, A. 2012, Monthly Notices of the
  Royal Astronomical Society, 419, 2133

\bibitem[{Hinshaw {et~al.}(2013)Hinshaw, Larson, Komatsu, Spergel, Bennett,
  Dunkley, Nolta, Halpern, Hill, Odegard, Page, Smith, Weiland, Gold, Jarosik,
  Kogut, Limon, Meyer, Tucker, Wollack, \& Wright}]{Hinshaw:2013zr}
Hinshaw, G., Larson, D., Komatsu, E., {et~al.} 2013, The Astrophysical Journal
  Supplement Series, 208, 19

\bibitem[{Huchra {et~al.}(1995)Huchra, Geller, \& Corwin}]{Huchra:1995bh}
Huchra, J.~P., Geller, M.~J., \& Corwin, Jr., H.~G. 1995, The Astrophysical
  Journal Supplement Series, 99, 391

\bibitem[{Johnson {et~al.}(2013)Johnson, Chen, \& Mulchaey}]{Johnson:2013lr}
Johnson, S.~D., Chen, H.-W., \& Mulchaey, J.~S. 2013, MNRAS

\bibitem[{Johnson {et~al.}(2015)Johnson, Chen, \& Mulchaey}]{Johnson:2015xy}
---. 2015, MNRAS, 449, 3263

\bibitem[{Johnson {et~al.}(2014)Johnson, Chen, Mulchaey, Tripp, Prochaska, \&
  Werk}]{Johnson:2014rt}
Johnson, S.~D., Chen, H.-W., Mulchaey, J.~S., {et~al.} 2014, MNRAS, 438, 3039

\bibitem[{Kacprzak {et~al.}(2012)Kacprzak, Churchill, Steidel, Spitler, \&
  Holtzman}]{Kacprzak:2012ul}
Kacprzak, G.~G., Churchill, C.~W., Steidel, C.~C., Spitler, L.~R., \& Holtzman,
  J.~A. 2012, MNRAS, 427, 3029

\bibitem[{Kere{\v s} {et~al.}(2005)Kere{\v s}, Katz, Weinberg, \&
  Dav{\'e}}]{Keres:2005lr}
Kere{\v s}, D., Katz, N., Weinberg, D.~H., \& Dav{\'e}, R. 2005, MNRAS, 363, 2

\bibitem[{Knobel {et~al.}(2015)Knobel, Lilly, Woo, \& Kova{\v
  c}}]{Knobel:2015uq}
Knobel, C., Lilly, S.~J., Woo, J., \& Kova{\v c}, K. 2015, ApJ, 800, 24

\bibitem[{Lehner {et~al.}(2013)Lehner, Howk, Tripp, Tumlinson, Prochaska,
  O'Meara, Thom, Werk, Fox, \& Ribaudo}]{Lehner:2013fj}
Lehner, N., Howk, J.~C., Tripp, T.~M., {et~al.} 2013, ApJ, 770, 138

\bibitem[{Liang \& Chen(2014)}]{Liang:2014kx}
Liang, C.~J., \& Chen, H.-W. 2014, MNRAS, 445, 2061

\bibitem[{Loveday {et~al.}(2012)Loveday, Norberg, Baldry, Driver, Hopkins,
  Peacock, Bamford, Liske, Bland-Hawthorn, Brough, Brown, Cameron, Conselice,
  Croom, Frenk, Gunawardhana, Hill, Jones, Kelvin, Kuijken, Nichol, Parkinson,
  Phillipps, Pimbblet, Popescu, Prescott, Robotham, Sharp, Sutherland, Taylor,
  Thomas, Tuffs, van Kampen, \& Wijesinghe}]{Loveday:2012jk}
Loveday, J., Norberg, P., Baldry, I.~K., {et~al.} 2012, MNRAS, 420, 1239

\bibitem[{Lupton {et~al.}(2004)Lupton, Blanton, Fekete, Hogg, O'Mullane,
  Szalay, \& Wherry}]{Lupton:2004yu}
Lupton, R., Blanton, M.~R., Fekete, G., {et~al.} 2004, Publications of the
  Astronomical Society of the Pacific, 116, 133

\bibitem[{Martin(2005)}]{Martin:2005qf}
Martin, C.~L. 2005, ApJ, 621, 227

\bibitem[{Martin {et~al.}(2015)Martin, Matuszewski, Morrissey, Neill, Moore,
  Cantalupo, Prochaska, \& Chang}]{Martin:2015kx}
Martin, D.~C., Matuszewski, M., Morrissey, P., {et~al.} 2015, Nature, 524, 192

\bibitem[{Meiring {et~al.}(2013)Meiring, Tripp, Werk, Howk, Jenkins, Prochaska,
  Lehner, \& Sembach}]{Meiring:2013fj}
Meiring, J.~D., Tripp, T.~M., Werk, J.~K., {et~al.} 2013, ApJ, 767, 49

\bibitem[{Morris {et~al.}(1993)Morris, Weymann, Dressler, McCarthy, Smith,
  Terrile, Giovanelli, \& Irwin}]{Morris:1993kq}
Morris, S.~L., Weymann, R.~J., Dressler, A., {et~al.} 1993, ApJ, 419, 524

\bibitem[{Moster {et~al.}(2013)Moster, Naab, \& White}]{Moster:2013lr}
Moster, B.~P., Naab, T., \& White, S. D.~M. 2013, MNRAS, 428, 3121

\bibitem[{Moster {et~al.}(2010)Moster, Somerville, Maulbetsch, van~den Bosch,
  Macci{\`o}, Naab, \& Oser}]{Moster:2010rt}
Moster, B.~P., Somerville, R.~S., Maulbetsch, C., {et~al.} 2010, ApJ, 710, 903

\bibitem[{Mould {et~al.}(2000)Mould, Huchra, Freedman, Kennicutt, Ferrarese,
  Ford, Gibson, Graham, Hughes, Illingworth, Kelson, Macri, Madore, Sakai,
  Sebo, Silbermann, \& Stetson}]{Mould:2000vn}
Mould, J.~R., Huchra, J.~P., Freedman, W.~L., {et~al.} 2000, ApJ, 529, 786

\bibitem[{Muldrew {et~al.}(2012)Muldrew, Croton, Skibba, Pearce, Ann, Baldry,
  Brough, Choi, Conselice, Cowan, Gallazzi, Gray, Gr{\"u}tzbauch, Li, Park,
  Pilipenko, Podgorzec, Robotham, Wilman, Yang, Zhang, \&
  Zibetti}]{Muldrew:2012qy}
Muldrew, S.~I., Croton, D.~J., Skibba, R.~A., {et~al.} 2012, Monthly Notices of
  the Royal Astronomical Society, 419, 2670

\bibitem[{Oppenheimer \& Dav{\'e}(2006)}]{Oppenheimer:2006uq}
Oppenheimer, B.~D., \& Dav{\'e}, R. 2006, MNRAS, 373, 1265

\bibitem[{Oppenheimer \& Dav{\'e}(2008)}]{Oppenheimer:2008mz}
---. 2008, MNRAS, 387, 577

\bibitem[{Oppenheimer {et~al.}(2012)Oppenheimer, Dav{\'e}, Katz, Kollmeier, \&
  Weinberg}]{Oppenheimer:2012qy}
Oppenheimer, B.~D., Dav{\'e}, R., Katz, N., Kollmeier, J.~A., \& Weinberg,
  D.~H. 2012, MNRAS, 420, 829

\bibitem[{Oppenheimer {et~al.}(2016)Oppenheimer, Crain, Schaye, Rahmati,
  Richings, Trayford, Tumlinson, Bower, Schaller, \&
  Theuns}]{Oppenheimer:2016lr}
Oppenheimer, B.~D., Crain, R.~A., Schaye, J., {et~al.} 2016, Monthly Notices of
  the Royal Astronomical Society, 460, 2157

\bibitem[{Osterbrock \& Ferland(2006)}]{Osterbrock:2006fk}
Osterbrock, D.~E., \& Ferland, G.~J. 2006, Astrophysics of gaseous nebulae and
  active galactic nuclei (Sausalito, CA: University Science Books)

\bibitem[{Prochaska {et~al.}(2011)Prochaska, Weiner, Chen, Mulchaey, \&
  Cooksey}]{Prochaska:2011yq}
Prochaska, J.~X., Weiner, B., Chen, H.-W., Mulchaey, J., \& Cooksey, K. 2011,
  ApJ, 740, 91

\bibitem[{Rahmati {et~al.}(2015)Rahmati, Schaye, Crain, Oppenheimer, Schaller,
  \& Theuns}]{Rahmati:2015lr}
Rahmati, A., Schaye, J., Crain, R.~A., {et~al.} 2015, ArXiv e-prints, 1511,
  1094

\bibitem[{Ribaudo {et~al.}(2011)Ribaudo, Lehner, Howk, Werk, Tripp, Prochaska,
  Meiring, \& Tumlinson}]{Ribaudo:2011qf}
Ribaudo, J., Lehner, N., Howk, J.~C., {et~al.} 2011, ApJ, 743, 207

\bibitem[{Rubin {et~al.}(2012)Rubin, Prochaska, Koo, \&
  Phillips}]{Rubin:2012fk}
Rubin, K. H.~R., Prochaska, J.~X., Koo, D.~C., \& Phillips, A.~C. 2012, The
  Astrophysical Journal Letters, 747, L26

\bibitem[{Rubin {et~al.}(2014)Rubin, Prochaska, Koo, Phillips, Martin, \&
  Winstrom}]{Rubin:2014fk}
Rubin, K. H.~R., Prochaska, J.~X., Koo, D.~C., {et~al.} 2014, ApJ, 794, 156

\bibitem[{Salim {et~al.}(2007)Salim, Rich, Charlot, Brinchmann, Johnson,
  Schiminovich, Seibert, Mallery, Heckman, Forster, Friedman, Martin,
  Morrissey, Neff, Small, Wyder, Bianchi, Donas, Lee, Madore, Milliard, Szalay,
  Welsh, \& Yi}]{Salim:2007lr}
Salim, S., Rich, R.~M., Charlot, S., {et~al.} 2007, The Astrophysical Journal
  Supplement Series, 173, 267

\bibitem[{Savage \& Sembach(1991)}]{Savage:1991vn}
Savage, B.~D., \& Sembach, K.~R. 1991, ApJ, 379, 245

\bibitem[{Schaye {et~al.}(2003)Schaye, Aguirre, Kim, Theuns, Rauch, \&
  Sargent}]{Schaye:2003rr}
Schaye, J., Aguirre, A., Kim, T.-S., {et~al.} 2003, ApJ, 596, 768

\bibitem[{Schechter(1976)}]{Schechter:1976yu}
Schechter, P. 1976, ApJ, 203, 297

\bibitem[{Schlegel {et~al.}(1998)Schlegel, Finkbeiner, \&
  Davis}]{Schlegel:1998kq}
Schlegel, D.~J., Finkbeiner, D.~P., \& Davis, M. 1998, ApJ, 500, 525

\bibitem[{Shen {et~al.}(2013)Shen, Madau, Conroy, Governato, \&
  Mayer}]{Shen:2013lr}
Shen, S., Madau, P., Conroy, C., Governato, F., \& Mayer, L. 2013, {ArXiv}
  e-prints, 1308, 4131

\bibitem[{Springel \& Hernquist(2003)}]{Springel:2003lr}
Springel, V., \& Hernquist, L. 2003, MNRAS, 339, 289

\bibitem[{Stocke {et~al.}(2013)Stocke, Keeney, Danforth, Shull, Froning, Green,
  Penton, \& Savage}]{Stocke:2013mz}
Stocke, J.~T., Keeney, B.~A., Danforth, C.~W., {et~al.} 2013, ApJ, 763, 148

\bibitem[{Strauss {et~al.}(2002)Strauss, Weinberg, Lupton, Narayanan, Annis,
  Bernardi, Blanton, Burles, Connolly, Dalcanton, Doi, Eisenstein, Frieman,
  Fukugita, Gunn, Ivezi{\'c}, Kent, Kim, Knapp, Kron, Munn, Newberg, Nichol,
  Okamura, Quinn, Richmond, Schlegel, Shimasaku, {SubbaRao}, Szalay,
  Vanden~Berk, Vogeley, Yanny, Yasuda, York, \& Zehavi}]{Strauss:2002fk}
Strauss, M.~A., Weinberg, D.~H., Lupton, R.~H., {et~al.} 2002, AJ, 124, 1810

\bibitem[{Taylor(2006)}]{Taylor:2006yu}
Taylor, M.~B. 2006, in , 666

\bibitem[{Tejos {et~al.}(2012)Tejos, Morris, Crighton, Theuns, Altay, \&
  Finn}]{Tejos:2012lr}
Tejos, N., Morris, S.~L., Crighton, N. H.~M., {et~al.} 2012, Monthly Notices of
  the Royal Astronomical Society, 425, 245

\bibitem[{Tremonti {et~al.}(2007)Tremonti, Moustakas, \&
  Diamond-Stanic}]{Tremonti:2007fk}
Tremonti, C.~A., Moustakas, J., \& Diamond-Stanic, A.~M. 2007, ApJL, 663, L77

\bibitem[{Tripp {et~al.}(2006)Tripp, Aracil, Bowen, \& Jenkins}]{Tripp:2006qy}
Tripp, T.~M., Aracil, B., Bowen, D.~V., \& Jenkins, E.~B. 2006, ApJL, 643, L77

\bibitem[{Tripp {et~al.}(1998)Tripp, Lu, \& Savage}]{Tripp:1998kq}
Tripp, T.~M., Lu, L., \& Savage, B.~D. 1998, The Astrophysical Journal, 508,
  200

\bibitem[{Tripp {et~al.}(2008)Tripp, Sembach, Bowen, Savage, Jenkins, Lehner,
  \& Richter}]{Tripp:2008lr}
Tripp, T.~M., Sembach, K.~R., Bowen, D.~V., {et~al.} 2008, ApJS, 177, 39

\bibitem[{Tripp {et~al.}(2011)Tripp, Meiring, Prochaska, Willmer, Howk, Werk,
  Jenkins, Bowen, Lehner, Sembach, Thom, \& Tumlinson}]{Tripp:2011wd}
Tripp, T.~M., Meiring, J.~D., Prochaska, J.~X., {et~al.} 2011, Science, 334,
  952

\bibitem[{Tumlinson {et~al.}(2011)Tumlinson, Thom, Werk, Prochaska, Tripp,
  Weinberg, Peeples, {O'Meara}, Oppenheimer, Meiring, Katz, Dav{\'e}, Ford, \&
  Sembach}]{Tumlinson:2011kx}
Tumlinson, J., Thom, C., Werk, J.~K., {et~al.} 2011, Science, 334, 948

\bibitem[{Tumlinson {et~al.}(2013)Tumlinson, Thom, Werk, Prochaska, Tripp,
  Katz, Dav{\'e}, Oppenheimer, Meiring, Ford, {O'Meara}, Peeples, Sembach, \&
  Weinberg}]{Tumlinson:2013cr}
---. 2013, ApJ, 777, 59

\bibitem[{Veilleux {et~al.}(2005)Veilleux, Cecil, \&
  Bland-Hawthorn}]{Veilleux:2005lq}
Veilleux, S., Cecil, G., \& Bland-Hawthorn, J. 2005, ARAA, 43, 769

\bibitem[{Verner {et~al.}(1994)Verner, Tytler, \& Barthel}]{Verner:1994fk}
Verner, D.~A., Tytler, D., \& Barthel, P.~D. 1994, ApJ, 430, 186

\bibitem[{Vogelsberger {et~al.}(2013)Vogelsberger, Genel, Sijacki, Torrey,
  Springel, \& Hernquist}]{Vogelsberger:2013qf}
Vogelsberger, M., Genel, S., Sijacki, D., {et~al.} 2013, Monthly Notices of the
  Royal Astronomical Society, 436, 3031

\bibitem[{Wakker {et~al.}(2015)Wakker, Hernandez, French, Kim, Oppenheimer, \&
  Savage}]{Wakker:2015rf}
Wakker, B.~P., Hernandez, A.~K., French, D.~M., {et~al.} 2015, The
  Astrophysical Journal, 814, 40

\bibitem[{Wakker \& Savage(2009)}]{Wakker:2009fr}
Wakker, B.~P., \& Savage, B.~D. 2009, ApJS, 182, 378

\bibitem[{Weinmann {et~al.}(2006)Weinmann, van~den Bosch, Yang, \&
  Mo}]{Weinmann:2006fj}
Weinmann, S.~M., van~den Bosch, F.~C., Yang, X., \& Mo, H.~J. 2006, MNRAS, 366,
  2

\bibitem[{Werk {et~al.}(2012)Werk, Prochaska, Thom, Tumlinson, Tripp,
  {O'Meara}, \& Meiring}]{Werk:2012qy}
Werk, J.~K., Prochaska, J.~X., Thom, C., {et~al.} 2012, ApJS, 198, 3

\bibitem[{Werk {et~al.}(2014)Werk, Prochaska, Tumlinson, Peeples, Tripp, Fox,
  Lehner, Thom, O'Meara, Ford, Bordoloi, Katz, Tejos, Oppenheimer, Dav{\'e}, \&
  Weinberg}]{Werk:2014kx}
Werk, J.~K., Prochaska, J.~X., Tumlinson, J., {et~al.} 2014, ApJ, 792, 8

\bibitem[{Willmer {et~al.}(2006)Willmer, Faber, Koo, Weiner, Newman, Coil,
  Connolly, Conroy, Cooper, Davis, Finkbeiner, Gerke, Guhathakurta, Harker,
  Kaiser, Kassin, Konidaris, Lin, Luppino, Madgwick, Noeske, Phillips, \&
  Yan}]{Willmer:2006lr}
Willmer, C. N.~A., Faber, S.~M., Koo, D.~C., {et~al.} 2006, The Astrophysical
  Journal, 647, 853

\bibitem[{Yang {et~al.}(2003)Yang, Mo, \& Bosch}]{Yang:2003yq}
Yang, X., Mo, H.~J., \& Bosch, F. C. v.~d. 2003, Monthly Notices of the Royal
  Astronomical Society, 339, 1057

\bibitem[{Yang {et~al.}(2007)Yang, Mo, van~den Bosch, Pasquali, Li, \&
  Barden}]{Yang:2007kl}
Yang, X., Mo, H.~J., van~den Bosch, F.~C., {et~al.} 2007, ApJ, 671, 153

\bibitem[{Yoon \& Putman(2013)}]{Yoon:2013kq}
Yoon, J.~H., \& Putman, M.~E. 2013, ApJL, 772, L29

\end{thebibliography}
\end{document}